\documentclass[longauth]{aa}  

\usepackage{graphicx}
\usepackage{txfonts}
\usepackage{xcolor}
\usepackage{multirow}
\usepackage{tabularx}
\usepackage{orcidlink}
\begin{document} 
    % TITLE IDEAS:
    % Gaia20fnr: A Non-Caustic-Crossing, Multi-Peak Microlensing Event with Orbital Motion Constrained by TESS Photometry
    % Gaia20fnr: A Non-Caustic-Crossing Binary-Lens Microlensing Event with Detectable Orbital Motion and TESS Photometry

    % Title Ideas:
    % - nearby
    % - bright
    % - TESS, \emph{Gaia}, WISE, Swift -- satellites
    % - off the galactic plane -- isolated
    % - orbital motion

   % \title{Gaia20fnr: A Non-Caustic-Crossing Binary-Lens Microlensing Event with Full Orbital Motion Revealed by Four Space Telescopes}
   \title{Gaia20fnr: A binary-lens microlensing event with full orbital motion revealed by four space telescopes}
   
   \authorrunning{Wicker et al.}
   \titlerunning{Gaia20fnr: Binary-lens microlensing event}
   \author{
    M. Wicker\inst{\ref{aff:UW}, \ref{aff:ING}}\orcidlink{0009-0006-7642-0943}
    ,
    {\L}. Wyrzykowski\inst{\ref{aff:UW}, \ref{aff:NCBJ}}\orcidlink{0000-0002-9658-6151}
    ,
    M. Hundertmark\inst{\ref{aff:ARI}}\orcidlink{0000-0003-0961-5231}
    ,
    K. A. Rybicki\inst{\ref{aff:UW}}\orcidlink{0000-0002-9326-9329}
    ,
    P. Zieli\'nski\inst{\ref{aff:Torun}}\orcidlink{0000-0001-6434-9429}
    ,
    E. Stonkut\.{e}\inst{\ref{aff:Vilnius}}\orcidlink{0000-0002-8028-8133}
    ,
    N. Ihanec\inst{\ref{aff:ING}}\orcidlink{0009-0003-7672-4806}
    ,
    M. Maskoli\=unas\inst{\ref{aff:Vilnius}}\orcidlink{0000-0003-3432-2393}
    ,
    E. Bachelet\inst{\ref{aff:CNRSBesancon}}\orcidlink{0000-0002-6578-5078}
    ,
    K. Kruszy\'{n}ska\inst{\ref{aff:LCO}}\orcidlink{0000-0002-2729-5369}
    ,
    M. Dominik\inst{\ref{aff:StAndrew}}\orcidlink{0000-0002-3202-0343}
    ,
    D. A. H. Buckley\inst{\ref{aff:saaoSA},\ref{aff:capetownSA}}\orcidlink{0000-0002-7004-9956}  
    ,
    I. Gezer\inst{\ref{aff:linzAustria}}
    ,
    M. Gromadzki\inst{\ref{aff:UW}}\orcidlink{0000-0002-1650-1518}
    ,
    P. Miko{\l}ajczyk\inst{\ref{aff:NCBJ},\ref{aff:Wroclaw}}\orcidlink{0000-0001-8916-8050}
    ,
    K. Kotysz\inst{\ref{aff:UW},\ref{aff:Wroclaw}}\orcidlink{0000-0003-4960-7463}
    ,
    J. Majumdar\inst{\ref{aff:UW}}\orcidlink{0009-0008-3960-1213}
    ,
    E. Pak\v{s}tien\.{e}\inst{\ref{aff:Vilnius}}\orcidlink{0000-0002-3326-2918}
    ,
    J. Zdanavi\v{c}ius\inst{\ref{aff:Vilnius}}
    ,
    V. \v{C}epas\inst{\ref{aff:Vilnius}}
    ,
    U. Jonauskait\.{e}\inst{\ref{aff:Vilnius}}\orcidlink{0009-0000-0293-8834} \\
    and \\ 
    % --- LCO-OMEGA ---
    V. Bozza\inst{\ref{aff:Salerno},\ref{aff:Napoli}}
    ,
    A. Cassan\inst{\ref{aff:Paris}}\orcidlink{0000-0003-4319-0628}
    ,
    R. Figuera Jaimes\inst{\ref{aff:MASchile},\ref{aff:PUCchile}}\orcidlink{0000-0003-3425-6605}
    ,
    M. Rabus\inst{\ref{aff:ConcepcionChile}}\orcidlink{0000-0003-2935-7196}
    ,
    P. Rota\inst{\ref{aff:Salerno},\ref{aff:Napoli}}\orcidlink{0000-0001-6161-1604}
    ,
    R. A. Street\inst{\ref{aff:LCO}}\orcidlink{0000-0001-6279-0552}
    ,
    Y. Tsapras\inst{\ref{aff:ARI}}\orcidlink{0000-0001-8411-351X}
    ,
    J. Wambsganss\inst{\ref{aff:ARI}}\orcidlink{0000-0001-5055-7390} \,
    (The OMEGA Key Project) \\
    and \\
    % --- BHTOM Observers ---
    S. Awiphan\inst{\ref{aff:Thailand}}
    \and
    S. M. Brincat\inst{\ref{aff:Flarestar}}
    \and
    Z. Budzik\inst{\ref{aff:UW}}
    \and
    J. W. Davidson Jr.\inst{\ref{aff:UniVirginia}}\orcidlink{0009-0007-1284-7240}
    \and
    R. Dymock\inst{\ref{aff:BritishAA}}
    \and
    C. Galdies\inst{\ref{aff:Malta}, \ref{aff:ZenithMalta}}
    \and
    V. Godunova\inst{\ref{aff:Ukraine}}\orcidlink{0000-0001-7668-7994}
    \and
    F.-J. Hambsch\inst{\ref{aff:VVSBelgium}, \ref{aff:SterneBerlin}}\orcidlink{0000-0003-0125-8700}
    \and
    M. Jab{\l}onska\inst{\ref{aff:UW}, \ref{aff:AustraliaStromlo}}
    \and
    T. Kvernadze\inst{\ref{aff:AbustamaniGeorgia}}
    \and
    M. Larma\inst{\ref{aff:NovaGorica}, \ref{aff:Bonn}}
    \and
    M. Makowska\inst{\ref{aff:Krakow}}
    \and
    Y. Markus\inst{\ref{aff:PavolKosice}, \ref{aff:Ukraine}}\orcidlink{0009-0006-1652-8581}
    \and
    J. Merc\inst{\ref{aff:Prague}, \ref{aff:IAC}}\orcidlink{0000-0001-6355-2468}
    \and
    O. Michniewicz\inst{\ref{aff:ZielonaGora}}
    \and
    M. Motylinski\inst{\ref{aff:Torun}}
    \and
    A. Popowicz\inst{\ref{aff:Gliwice}}
    \and
    M. Radziwonowicz\inst{\ref{aff:UW}}
    \and
    D. Reichart\inst{\ref{aff:NorthCarolina}}
    \and
    A.O. Simon\inst{\ref{aff:TarasKyiv}, \ref{aff:JunioarAcademy}}\orcidlink{0000-0003-0404-5559}
    \and
    P. Trzcionkowski\inst{\ref{aff:UW}}
    \and
    M. Zejmo\inst{\ref{aff:ZielonaGora}}
    \and
    S. Zola\inst{\ref{aff:JagiellonianU}} \,
    (BHTOM Observers) \\
    % \and
    }
   \institute{
    % 1.
    \label{aff:UW}Astronomical Observatory, University of Warsaw, Al. Ujazdowskie 4, 00-478 Warsaw, Poland \\
    \email{mwicker@astrouw.edu.pl}
    \and
    % 2.
    \label{aff:ING}Isaac Newton Group of Telescopes, Apartado 321, E-38700 Santa Cruz de La Palma, Spain
    \and
    % 3.
    \label{aff:NCBJ}Astrophysics Division, National Centre for Nuclear Research, Pasteura 7, 02-093, Warsaw, Poland
    \and
    % 6.
    \label{aff:ARI}Astronomisches Rechen-Institut, Zentrum f{\"u}r Astronomie der Universit{\"a}t Heidelberg (ZAH), 69120 Heidelberg, Germany
    \and
    % 5.
    \label{aff:Torun}Institute of Astronomy, Faculty of Physics, Astronomy and Informatics, Nicolaus Copernicus University, Grudzi\k{a}dzka 5, 87-100 Toru{\'n}, Poland
    \and
    % 4.
    \label{aff:Vilnius}Vilnius University, Faculty of Physics, Institute of Theoretical Physics and Astronomy, Sauletekio av. 3, 10257 Vilnius, Lithuania
    \and
    % 8.
    \label{aff:CNRSBesancon}Universit{\'e} Marie et Louis Pasteur, CNRS, Institut UTINAM UMR 6213, Besan\c{c}on, France
    \and
    % 7.
    \label{aff:LCO}Las Cumbres Observatory, 6740 Cortona Drive, Suite 102, 93117 Goleta, CA, USA
    \and
    % 9.
    \label{aff:StAndrew}SUPA, School of Physics \& Astronomy, University of St Andrews, North Haugh, St Andrews KY16 9SS, UK
    \and
    \label{aff:saaoSA}South African Astronomical Observatory, PO Box 9, Observatory 7935, Cape Town, South Africa
    \and
    \label{aff:capetownSA}Department of Astronomy, University of Cape Town, Private Bag X3, Rondebosch 7701, South Africa
    \and
    \label{aff:linzAustria}Institute for Machine Learning, Johannes Kepler University Linz, Altenberger Straße 69, 694040 Linz, Austria
    \and
    % 10.
    \label{aff:Wroclaw}Astronomical Institute, University of Wroc{\l}aw, ul. Miko{\l}aja Kopernika 11, 51-622 Wroc{\l}aw, Poland
    \and
    % LCO-OMEGA
    \label{aff:Salerno}Dipartimento di Fisica "E.R. Caianiello", Universit{\'a} di Salerno, Via Giovanni Paolo II 132, I-84084 Fisciano, Italy
    \and
    \label{aff:Napoli}Istituto Nazionale di Fisica Nucleare, Sezione di Napoli, Via Cintia, I-80126 Napoli, Italy
    \and
    \label{aff:Paris}Institut d’Astrophysique de Paris, Sorbonne Universit{\'e}, CNRS, UMR 7095, 98 bis bd Arago, F-75014 Paris, France
    \and
    \label{aff:MASchile}Millennium Institute of Astrophysics MAS, Nuncio Monsenor Sotero Sanz 100, Of. 104, Providencia, Santiago, Chile
    \and
    \label{aff:PUCchile}Instituto de Astrof{\'i}sica, Facultad de F{\'i}sica, Pontificia Universidad Cat{\'o}lica de Chile, Av. Vicu{\~n}a Mackenna 4860, 7820436 Macul, Santiago, Chile
    \and
    \label{aff:ConcepcionChile}Departamento de Matem{\'a}tica y F{\'i}sica Aplicadas, Facultad de Ingenier{\'i}a, Universidad Cat{\'o}lica de la Sant{\'i}sima Concepci{\'o}n, Alonso de Rivera 2850, Concepci{\'o}n, Chile
    \and
    % BHTOM Observatories
    \label{aff:Thailand}National Astronomical Research Institute of Thailand (Public Organization), 260 Moo 4, Donkaew, Mae Rim, Chiang Mai 50180, Thailand
    \and
    \label{aff:Flarestar}Flarestar Observatory, FL5 Ent.B, Silver Jubilee Apt, George Tayar Street, San Gwann, SGN 3160, Malta
    \and
    \label{aff:UniVirginia}Department of Astronomy, University of Virginia, 530 McCormick Road, Charlottesville, VA 22904, USA
    \and
    \label{aff:BritishAA}British Astronomical Association, PO Box 702, Tonbridge, TN9 9TX England
    \and
    \label{aff:Malta}Institute of Earth Systems, University of Malta, Msida MSD 2080, Malta
    \and
    \label{aff:ZenithMalta}Znith Astronomy Observatory, 31, Armonie, E. Bradford Street, Naxxar 2217, Malta
    \and
    \label{aff:Ukraine}International Center for Astronomical and Medico-Ecological Research (ICAMER), National Academy of Sciences of Ukraine, 27 Acad.\ Zabolotnoho Str., Kyiv, 03143, Ukraine
    \and
    \label{aff:VVSBelgium}Vereniging Voor Sterrenkunde (VVS), Zeeweg 96, B-8200 Brugge, Belgium
    \and
    \label{aff:SterneBerlin}Bundesdeutsche Arbeitsgemeinschaft f{\"u}r Ver{\"a}nderliche Sterne, Munsterdamm 90, D-12169 Berlin, Germany
    \and
    \label{aff:AustraliaStromlo}Research School of Astronomy and Astrophysics, Australian National University, Mount Stromlo Observatory, Cotter Road Weston Creek, ACT 2611, Australia
    \and
    \label{aff:AbustamaniGeorgia}E. Kharadze Georgian National Astrophysical Observatory, 0301 Abastumani, Georgia
    \and
    \label{aff:NovaGorica}Center for Astrophysics and Cosmology, University of Nova Gorica, Vipavska 11c, 5270 Ajdov{\v{s}}cina, Slovenia
    \and
    \label{aff:Bonn}Argelander-Institut f{\"u}r Astronomie, Universit{\"a}t Bonn, Auf dem H{\"u}gel 71, 53121 Bonn, Germany
    \and
    \label{aff:Krakow}Instytut Fizyki im.\ Mariana Smoluchowskiego, Uniwersytet Jagiello{\'n}ski, ul.\ prof.\ St.\ {\L}ojasiewicza 11, 30-348 Krak\'ow, Poland
    \and
    \label{aff:PavolKosice}Institute of Physics, Faculty of Science, Pavol Jozef Šaf{\'a}rik University in Ko{\v{s}}ice, Park Angelinum 9, 040 01 Ko{\v{s}}ice, Slovakia
    \and
    \label{aff:Prague}Astronomical Institute of Charles University, V Hole{\v{s}}ovi{\v{c}}k{\'a}ch 2, Prague 18000, Czech Republic
    \and
    \label{aff:IAC}Instituto de Astrofísica de Canarias, Calle V{\'i}a L{\'a}ctea, s/n E-38205 La Laguna, Tenerife, Spain
    \and
    \label{aff:ZielonaGora}Janusz Gil Institute of Astronomy, University of Zielona G{\'o}ra, Szafrana 2, 65–516 Zielona Góra, Poland
    \and
    \label{aff:Gliwice}Faculty of Automatic Control, Electronics and Computer Science, Silesian University of Technology, Akademicka 16, 44-100 Gliwice, Poland
    \and
    \label{aff:NorthCarolina}Department of Physics and Astronomy, University of North Carolina at Chapel Hill, Chapel Hill, NC 27599, USA
    % \and
    % \label{aff:INAF}INAF - Osservatorio Astronomico di Roma, Via di Frascati 33, I-00078 Monteporzio Catone, Italy
    \and
    \label{aff:TarasKyiv}Astronomy and Space Physics Department, Taras Shevchenko National University of Kyiv, 4 Glushkova ave., Kyiv, 03022, Ukraine
    \and
    \label{aff:JunioarAcademy}National Center "Junior Academy of Sciences of Ukraine", 38-44 Dehtiarivska St., Kyiv, 04119, Ukraine
    \and
    \label{aff:JagiellonianU}Astronomical Observatory, Jagiellonian University, ul. Orla 171, 30-244 Kraków, Poland
    % \and
    } 
   \date{Received December AB, WXYZ; accepted XXX AB, WXYZ}

   \abstract{
    % Context.
    {The microlensing event Gaia20fnr is a long-duration, non-caustic-crossing binary-lens event located at high Galactic latitude. Triggered by a photometric rise detected by the \emph{Gaia} space mission, the event was followed up with observations from multiple ground-based facilities and observed by four space telescopes: \emph{Gaia}, NEOWISE, Swift, and TESS.}
    % Aims.
    {We characterize the Gaia20fnr microlensing system by determining the physical and orbital properties of the binary lens, the nature of the luminous source, and the kinematics of both the source and the lens.}
    % Methods.
    {We employed a binary-lens microlensing model including full Keplerian orbital motion and annual microlens parallax to fit the photometric data.}
    % Results.
    {The event is best explained by a K2 giant source at $D_{\rm{S}} = 3.10 \pm 0.10\,\mathrm{kpc}$ lensed by a stellar binary composed of $M_{\rm{L},1} = 0.46 \pm 0.06\,M_{\odot}$ and $M_{\rm{L},2} = 0.52 \pm 0.06\,M_{\odot}$ at a distance of $D_{\rm{L}} = 0.54 \pm 0.05\,\mathrm{kpc}$. The light curve exhibits strong signatures of orbital motion and requires a full Keplerian model with a period of $P = 0.67 \pm 0.04\,\mathrm{yr}$ and a radial-velocity semi-amplitude of $K_1 = 16.9 \pm 0.9\,\mathrm{km\,s^{-1}}$ to be explained.}
    % Conclusions.
    {Gaia20fnr is one of the few microlensing events for which a complete Keplerian binary-lens solution has been derived. The model can be tested with follow-up radial-velocity and high-resolution imaging observations as well as forthcoming \emph{Gaia} DR4 and DR5 astrometric time-series data. Its long duration, multi-peak structure, and extensive coverage from both space- and ground-based facilities make it a benchmark for studying faint nearby low-mass binaries through microlensing.}
    }
   

   \maketitle
%

% Edita et al. - checking the archives with spectra and parameters
% -> not found any crossmatches in Galah, DESI DR1, APOGEE DR17, LAMOST LRS DR9.}

%-------------------------------------------------------------------
\section{Introduction}
Binary star systems are essential to understanding stellar astrophysics. They provide measurements of stellar masses and luminosities, which strongly constrain the mass–luminosity relation and models of stellar evolution \citep{Paczynski1971Binaries, boffin2024importancebinarystars}. Studies of stellar binaries have typically relied on direct measurements of light emitted by the binary system, such as radial-velocities, to determine the mass and interaction between bodies \citep{popperBinaries, TorresBinaries}. These traditional techniques are restricted to a small and systematically biased subset of binary systems, as precise dynamical solutions require bright systems with well-measured radial-velocity curves and, ideally, eclipses. As a result, samples with accurately known masses and radii remain small and strongly biased toward nearby, relatively unevolved, and non-interacting systems \citep{TorresBinaries, Morales_2022}. Population studies further show that a large fraction of binaries, especially long-period systems, low-mass companions, and systems containing compact remnants, remain essentially unconstrained by direct light and radial-velocity measurements. This leads to severe selection effects in inferred binary demographics \citep{MoeBinaries, CasaresBinaries}. Alternative detection techniques are essential because these selection effects leave large portions of binary-parameter space unexplored. 

In 1986, Bohdan Paczy{\'n}ski proposed gravitational microlensing as a method to detect compact dark matter objects, such as Massive Compact Halo Objects (MACHOs), in the Galactic Halo by observing their lensing effects on background stars \citep{PaczynskiMicrolensing86,AlcockMacho}. Since then, gravitational microlensing has proven to be an effective method for detecting isolated and otherwise hard-to-find objects such as planetary systems \citep[e.g.][]{Bond2004, BeaulieuExopl, BennettCircumbin} free-floating or wide-orbit planets \citep[e.g.][]{Mroz2018FFP,Mroz2019FFP, poleskiWideOrbit, newFFPscience}, and stellar-mass black holes \citep{SahuBlackHole,LamDarkRem2022}. These studies have provided estimates and constraints for stellar, planetary, and dark-remnant populations \citep[e.g.][]{Wyrzykowski2016Remnants, Sweeney2024, Kruszynska2024, Abrams2025, Kaczmarek2025, Mroz2025opticalDepth} as well as explanations for complex, multi-peak microlensing light curves that are difficult to explain by simpler lens models or other astrophysical sources of variability \citep[e.g.][]{Udalski4Peak, Wyrzykowski16aye}. Gravitational microlensing has proven particularly powerful for discovering and characterizing binary systems that are faint, distant, or exhibit significant orbital motion \citep[e.g.][]{Skowron2011,Penny2011}.

Large microlensing surveys (e.g. OGLE \citep{OGLE2}; MOA \citep{MOA1}; KMT \citep{KMT1}; PRIME \citep{PrimePaper}) typically target regions of high stellar density, such as the Galactic bulge and Magellanic Clouds, to maximise the event rate. While these surveys rely on monitoring dense stellar fields from the ground, the \emph{Gaia} mission provides an entirely different and complementary approach to microlensing discovery. Operating from space, \emph{Gaia} conducted an all-sky high-precision astrometric and photometric survey that observed the entire sky. As a result, \emph{Gaia} was sensitive to microlensing events across the entire sky, including regions that are sparsely or not at all monitored by traditional ground-based programs. This makes \emph{Gaia} uniquely valuable for uncovering objects acting as gravitational lenses that could otherwise be underrepresented, such as nearby lenses and high proper-motion objects. The first catalogue of \emph{Gaia} microlensing events was presented in \citet{WyrzykowskiDR3Microlensing} as part of \emph{Gaia} Data Release 3 (DR3).

In order to make use of transients observed by \emph{Gaia} and to enable timely follow-up observations, the \emph{Gaia} Science Alerts (GSA; \citet{HodgkinGSA}) system identified transient brightness variations in real time. This enabled the coordination of fast follow-up observations for a variety of transients. Follow-up observations, both photometric and spectroscopic, have produced results of GSA for a variety of sources, such as eclipsing AM CVn-systems \citep[e.g.][]{Campbell_2015}, superluminous supernovae \citep[e.g.][]{Kangas_SN, Nicholl_2017}, nuclear transients \citep[e.g.][]{KostrzewaTransients, PessiTransient}, cataclysmic and symbiotic variables \citep[e.g.][]{Mercsymbitioc2020, Mistry_2022, Mercsymbitioc2024}, EM-counterparts to gravitational-wave observations \citep[e.g][]{KostrzewaEM}, and gravitational microlensing events \citep[e.g.][]{Kris19bld, Wyrzykowski16aye, Kruszynska2022Gaia18cbf, PylypenkoML}.

\emph{Gaia}’s forthcoming Data Release 4 (DR4), expected in December 2026, will include full astrometric time-series data for all sources with epoch-by-epoch centroid positions rather than only the five-parameter average solutions of previous releases. Astrometric microlensing modelling will allow for direct measurement of the lens mass and distance \citep{Jankovi2025}. The joint astrometric and photometric data will be key in breaking the degeneracies inherent in photometric-only microlensing observations and will challenge current results derived from photometric models.

Population studies show that approximately $5$–$10\,\%$ of microlensing events exhibit signatures of binary lenses \citep{MaoPaczynski, Udalski2000, Dominik_2006, oliveira2025}. Unlike single-lens events, which produce a simple symmetric light curve, binary lenses can generate complex caustic structures that depend sensitively on the mass ratio and projected separation of the components \citep{Bozza_2001, Rattenbury_2002, Cassan_2008}. These caustic networks produce sharp, high-magnification features that can tightly constrain lens geometry when the photometric coverage densely samples caustic crossings or cusp approaches, where the magnification pattern is particularly sensitive to binary geometry \citep{JaroszyMao,Skowron2007}. In addition to the standard binary-lens parameters, higher-order effects are often required to obtain physical solutions. These include microlens parallax, caused by the observer’s motion around the Sun or by observations from separated observatories, and finite-source effects associated with limb-darkened source profiles. Together, these effects break degeneracies and permit measurements of the lens mass and distance \citep{WittMaoFiniteSource, GouldJerk}. Incorporating these effects increases model complexity but is essential for converting microlensing observables into robust physical parameters \citep{Bennett_2007, Dong2009}. For binary-lens microlensing events, higher-order effects from orbital motion can produce time-varying caustic structures that substantially alter the resulting light curve.

Wide binaries of low-mass (solar-type) stars typically have orbital periods of order a few hundred years \citep{Raghavan2010}, far exceeding the Einstein crossing time $t_E$ of most microlensing events. In contrast, close binaries, such as massive-star pairs or post-common-envelope systems, can have orbital periods $P$ comparable to or shorter than $t_E$. In general, for binary lenses with $P \gg t_E$, orbital motion can be neglected in microlensing models \citep{Penny2011, DiStefano}. However, when $P \lesssim t_E$, orbital evolution can significantly alter the projected separation and orientation of the lens components, thereby modifying the caustic morphology and producing measurable deviations in the light curve relative to static binary-lens models, which can no longer provide satisfactory fits \citep{Albrow_2000,Nucita2014,Guo_2015}. Notably, \citet{Shin_2012} found that systems with $P \sim 1$--$3\,\mathrm{yr}$ could not be satisfactorily modelled without including orbital motion, even though their periods exceeded typical $t_E$ values. Long event timescales and the proximity to resonant caustic configurations render light curves sensitive to small changes in the projected binary geometry.

Although rotating binary stars increase the complexity of the microlensing model, the corresponding Keplerian orbital parameters (period, velocity amplitude, eccentricity, phase, and position of periapsis) provide an independent means of testing the model via radial-velocity measurements. This possibility was demonstrated for the \emph{OGLE-2009-BLG-020} event, where the prediction of \citet{Skowron2011} was subsequently verified through radial-velocity measurements by \citet{Yee2016}.

Here we present a detailed analysis of a unique microlensing event characterised by significant binary orbital motion and no caustic-crossing features, Gaia20fnr. This event provides a rare opportunity to constrain the orbital parameters of the binary lens system, characterised by its five Keplerian parameters. The photometric event lasted for over a year, with multiple distinct features in the light curve. Apart from a short period before the maximum magnification where the event was unobservable because of the proximity of the Sun to the line of sight of the microlensing event, the event was observed for the entirety of the source magnification, with ample photometric data at the baseline before and after the source magnification. Due to an early alert from \emph{Gaia}, a large network of ground-based telescopes coordinated by the Black Hole Target and Observation Manager (BHTOM\footnote{\url{https://bhtom.space}}) \citep{BHTOM1, BHTOM2} was able to observe the event in a variety of filters. BHTOM facilitates coordinated observations and the processing of photometric time-series data, building on the open-source TOM Toolkit developed at Las Cumbres Observatory \citep[LCO;][]{StreetTOM, StreetTOM2024}.
The high cadence of observations, especially during each of the three magnification periods, enabled us to find a unique solution with well-constrained parameters for a single source and a rotating binary lens. The detailed modelling presented here aims not only to impose lens mass estimates but also to provide robust parameters that can be independently verified by subsequent radial-velocity and astrometric data, mirroring the approaches demonstrated by \citet{Skowron2011} and \citet{Wyrzykowski16aye}.

The structure of this paper is as follows: in Sections 2 and 3, we detail the \emph{Gaia} alert, the photometric data, and the spectroscopic observations. Sections 4 and 5 outline microlensing modelling, particularly emphasising the incorporation of orbital motion to model the data and the physical parameters of the source and lens. The results are discussed in Section 6.
%--------------------------------------------------------------------
\section{Discovery and photometric follow-up}
\begin{figure*}[t]
    \centering
    \includegraphics[width=\textwidth]{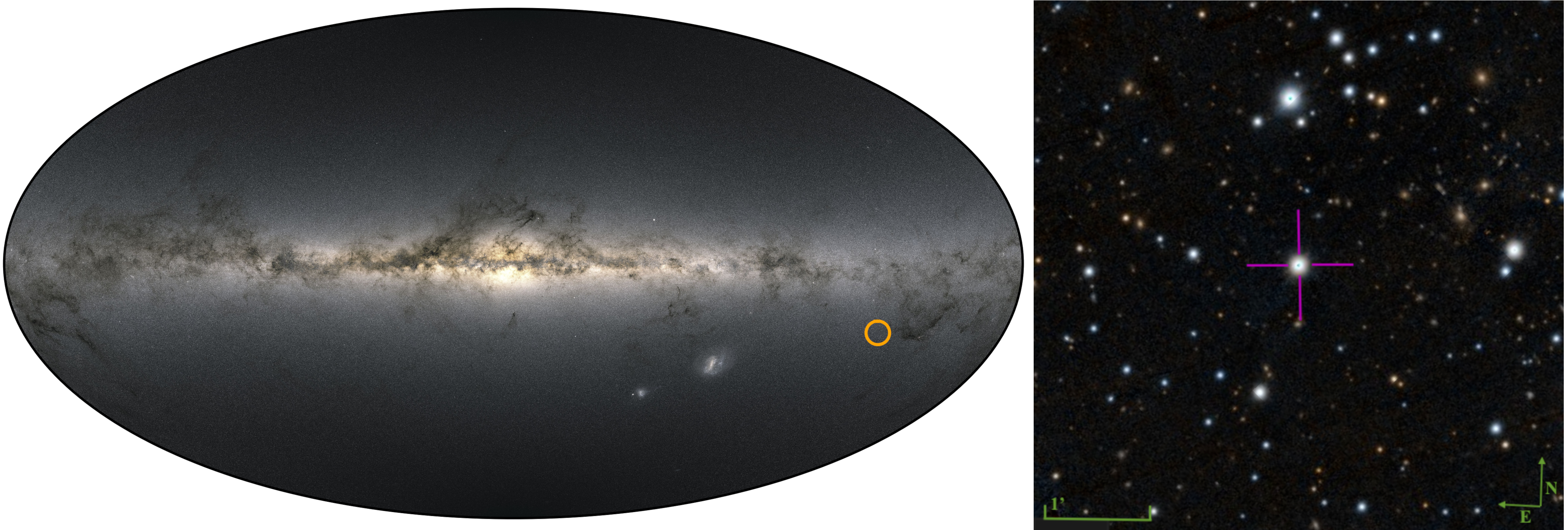}
    \caption{Sky map view of the Milky Way with the location of Gaia20fnr marked with an orange circle. This illustrates the unique location of the microlensing event. The figure was made with MW-Plot\textsuperscript{a} and \emph{Gaia} EDR3 data \citep{GaiaEDR3}. On the right, a finding chart made from Pan-STARRS1 \citep{Panstarrs1} data via the \emph{ALADIN} Tool \citep{aladinSky}. }
    \label{fig:milkywayViewplot}
\end{figure*}

The Gaia20fnr microlensing event (IAU designation AT2020ably in the Transient Name Server), located at equatorial coordinates $\left( \alpha, \delta \right)_{J2000} = \left( 06^\mathrm{h} 01^\mathrm{m} 04.08^\mathrm{s} , -18 ^\circ 58' 03.72'' \right)$, was identified as a transient event by GSA\footnote{\url{https://gsaweb.ast.cam.ac.uk/alerts/alert/Gaia20fnr/}} on 2 December 2020 \citep{2020TNS20fnr}. The Galactic coordinates of the event are $\left( l, b \right) = \left(224.878^\circ, -19.372^\circ \right)$, such that Gaia20fnr is in a region of the Milky Way with low stellar density and far from the Galactic Bulge and Disk. The finding chart and sky location of Gaia20fnr are shown in Fig.~\ref{fig:milkywayViewplot}\footnote{MW-Plot: \url{https://milkyway-plot.readthedocs.io/en/stable/}}.

The \emph{Gaia} DR3 catalogue \citep{GaiaDR3} source parameters for the alerted \emph{Gaia}  object are listed in Table \ref{tab:Gaia5param}. The source also has a counterpart in the Two Micron All Sky Survey \citep[2MASS;][]{2mass} as 2MASS 06010409-1858038. The alert was issued following a 0.26 mag rise in the \emph{Gaia} G-band relative to five years of pre-alert observations, during which the source had a stable mean magnitude of 13.17 mag ($\sigma = 0.02 \,\mathrm{mag}$). Equal magnification in the \emph{Gaia} blue (BP) and red (RP) bands indicated an achromatic brightening consistent with microlensing. This motivated an extensive follow-up campaign.

% Include a table of \emph{Gaia} DR3 5-param solution parameters
\begin{table}[h!]
\caption{\emph{Gaia} DR3 parameters for the source of the Gaia20fnr microlensing event.}
\centering
\begin{tabular}{c | c}
\emph{Gaia} Source ID & 2990431491637998848 \\
\hline
(RA, Dec) $[\mathrm{deg}]$ & ($90.27, \, -18.97$) \\
(l, b) $[\mathrm{deg}]$ & ($224.88, \, -19.37$) \\
$\varpi$ $[\mathrm{mas}]$ & $0.29 \, \pm \, 0.01$ \\
$\mu_{\alpha*}$ $[\mathrm{mas\,yr^{-1}}]$ & $-2.25 \, \pm \, 0.01$ \\
$\mu_{\delta}$ $[\mathrm{mas\,yr^{-1}}]$ & $2.72 \, \pm \, 0.01$ \\
$m_G$ [mag] & $13.106 \, \pm 0.003$ \\
$m_{BP}$ [mag] & $13.646 \, \pm 0.003$ \\
$m_{RP}$ [mag] & $12.413 \, \pm 0.004$ \\
RUWE & $1.026$ \\
% \hline
\end{tabular}
\label{tab:Gaia5param}
\end{table}

The source has historically been observed by a multitude of surveys and in a variety of filters. Such data obtained before, during, and after the magnification of the source are crucial for modelling and accurately identifying source and blend fluxes, as they provide a baseline measure of the unmagnified source flux. Additionally, in order to rule out causes besides microlensing and to determine potential periodic variations/oscillations, historic and continuous observations are vital. The Catalina Real-Time Transient Survey \citep{CRTS1, CRTS2} observed the source for a 10-year period from 2005 and shows no significant variability. The Galaxy Evolution Explorer 
\citep[GALEX;][]{GALEX} space telescope observed the source in 2007 with its far ultraviolet (FUV) and near ultraviolet (NUV) filters. The Palomar Transient Factory (PTF) \citep{PTF} observed the source between 2009 and 2012 in the R-band. The Panoramic Survey Telescope and Rapid Response System (PS1) \citep{chambers2019panstarrs1surveys} provided data recorded between 2011 and 2013 in the g-, r-, i-, and z-bands. Since November 2013, the All-Sky Automated Survey for Supernovae (ASAS-SN) \citep{ASASSN} has been observing the source and continued to observe it for the entire microlensing event. The relatively high cadence of two to five days in the V- and g-bands results in extensive observational coverage. The Zwicky Transient Facility (ZTF) \citep{ZTFBellm} observed the target in the r- and g-bands since November 2018. The ZTF data cover the baseline before magnification, the first peak, the dimming phase following the main peak, and the post-event baseline of the photometrically observable microlensing event. The Asteroid Terrestrial-impact Last Alert System (ATLAS) \citep{ATLAS} has been monitoring the source since 2015, providing coverage of the microlensing event with a cadence of roughly four days in the cyan and orange bands.

In addition to \emph{Gaia}, this event was monitored by three other space-based observatories. Together, these facilities provided coverage from the infrared to the ultraviolet, making the dataset unique for a microlensing event. The Transiting Exoplanet Survey Satellite (TESS) \citep{TESS} observed the source during three sectors beginning in December 2018, December 2020, and December 2024, with respective time spans of approximately 22, 26, and 27 days. Complementary infrared observations have been obtained since March 2010 by NEOWISE \citep{NEOWISE}, while ultraviolet monitoring has been provided since April 2021 by the Swift UV/Optical Telescope \citep{Swfittelescope}.

Following the alert from \emph{Gaia}, photometric follow-up observations were performed by a large number of telescopes in a variety of visible bands. The observations for this event\footnote{\url{https://bh-tom2.astrolabs.pl/public/targets/Gaia20fnr}} were collected, and more recently also coordinated, by the BHTOM network of telescopes. Twenty three different ground-based telescopes provided follow-up observations for five Johnson-Cousins (JC; \citet{Bessell_JC}) and five Sloan Digital Sky Survey (SDSS) \citep{SDSS_filts} filters.

\begin{figure*}[t]
\centering
\includegraphics[width=\textwidth]{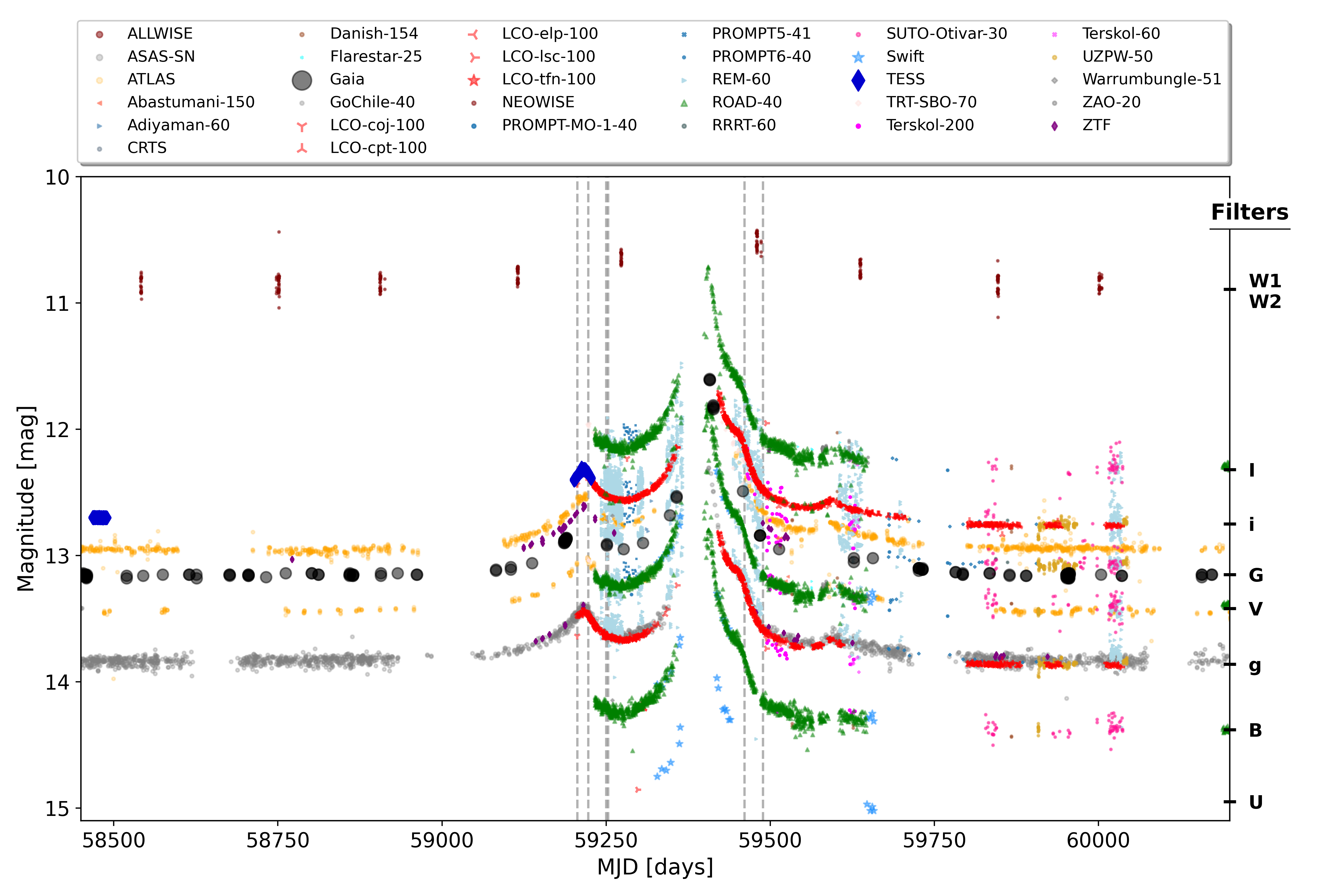}
\caption{Photometric observations of Gaia20fnr combining space-based observations, survey data, and follow-up observations gathered by BHTOM. The Swift UV data in the UVW1 and UVW2 filter are not shown for visibility reasons. Their data are at the same timestamps as the Swift U-band data, but with 1.6 mag and 2.8 mag higher magnitude values for UVW1 and UVW2, respectively. The dashed grey lines mark the times at which spectroscopic observations were obtained.}
\label{fig:allPhotdata}
\end{figure*}

The collection of photometric data, shown for each telescope, is in Fig.~\ref{fig:allPhotdata}. The ground-based telescopes that provided photometric follow-up observations are detailed in Table~\ref{tab:telescopeInfo} and details of all photometric observations are listed in Table~\ref{tab:allData}. 

\subsection{\emph{Gaia} photometry}
An alert was issued for the Gaia20fnr microlensing event in 2020, with the closest approach between the lens and the source occurring in 2021, and the return to the baseline at the end of 2022. \emph{Gaia} DR4 will contain data from observations between July 2014 and January 2020, which only covers the first peak of the photometric microlensing event. Only \emph{Gaia} Data Release 5 (DR5), which is not expected before the end of 2030, will contain the full light curve and also the full astrometric time series. Full and up-to-date photometric G-band observations by the \emph{Gaia} telescope are published by the GSA system. These, however do not provide magnitude errors for published events. In order to estimate these error bars, we followed the method of \citet{Kruszynska2022Gaia18cbf}. 

\subsection{Ground-based follow-up}
The Gaia20fnr alert came more than 200 days before the closest approach between the source and the lens. This has enabled an extensive photometric follow-up campaign to collect data for the majority of the event, covering almost all features in their entirety. The very first photometric follow-up observations started 20 days after the alert was issued, on the 22 December 2020, by the OMEGA Project, using the LCO Global Telescope Network of 1-m telescopes \citep{LCOBrown}. LCO data in g- and i-bands were obtained using the 1-m Sinistro-camera telescopes located in Australia, Chile, South Africa, Tenerife, and the USA.

Because of the event’s position on the sky relative to the Sun, it was unobservable from Earth for a period leading up to the main peak. Nevertheless, observations from the ROAD observatory \citep{RoadObservatory} pushed the observing limits and captured much of the peak, which was critical for constraining the model parameters and for potentially detecting finite-source effects. On 7 October 2021, almost a year after the alert of the transient event, the object reached its peak magnitude of $m_\mathrm{V} = 11.8 \, \mathrm{mag}$. The third and smallest brightening phase in January of 2022 was extensively observed and was crucial in constraining the orbital motion parameters despite its relatively small magnification of 1.2 times the baseline source flux (compared to 1.5 for the first and 4.35 for the second peak).

All of the ground-based photometric observations were processed using the BHTOM tool, an automated system designed for managing the targets and processing of the time-domain data. Bias-, dark-, and flat-field-corrected images were analysed to extract instrumental photometry for all detected stars using the package CCDPhot built into BHTOM. The resulting photometry was subsequently standardised \citep{zielinski2019, zielinski2020automaticprocessingccdimages} using \emph{Gaia} Synthetic Photometric (SP) magnitudes \citep{MontegriffoGaiaSPC}.

\subsection{Swift observations}
The Neil Gehrels Swift Observatory \citep{Swfittelescope} is a multi-wavelength space telescope designed for rapid-response observations of transient phenomena. In addition to its X-ray capabilities, Swift’s Ultraviolet/Optical Telescope (UVOT) provides sensitive imaging in the near-ultraviolet and optical bands. The UV coverage enabled the investigation of potential chromatic effects, blending contributions, or signatures associated with the source or lens that may be challenging to detect at longer wavelengths.

We triggered Swift through its Target-of-Opportunity programme and obtained observations for eighteen epochs with both the UVOT and X-ray Telescope (XRT), operating in the photon-counting mode. The UVOT light curves were extracted using a $5''$ aperture and a $20''$ background region. The data were reduced using the {\sc HEASOFT}\,v6.36 software package and the \texttt{uvotsource} tool \citep{heasoft}, and the resulting photometry was subsequently uploaded to BHTOM. The Swift/XRT observations were processed using the standard {\sc xrtpipeline} procedures. No X-ray source was detected in any epoch. The XRT and UVOT coverage spans times before, during, and after the main magnification region, including several high-magnification observations.

\subsection{TESS observations}
The Transiting Exoplanet Survey Satellite \citep{TESS} is a space-based mission designed to perform high-precision, uninterrupted photometry over wide fields for transiting exoplanets. Unlike ground-based surveys, TESS provides nearly continuous monitoring over $\sim 27$-day sectors with minimal gaps and without the limitations imposed by weather, seeing, or diurnal cycles. Although optimised for exoplanet detection, TESS has proven valuable for time-domain astrophysics more broadly, offering high-cadence light curves that are sensitive to transient events like microlensing.

TESS observed the source (TIC 123793270) in 2018, 2020, and 2024 during Sectors 06, 33, and 87, respectively. Full-frame image (FFI) photometry was obtained with cadences of 30 minutes (Sector 06), 10 minutes (Sector 33), and 200 seconds (Sector 87). The TESS light curve covers a baseline period in 2018/2019, a period during the primary magnification region (from December 2020 until January 2021), and a period when the event had returned to its baseline magnitude (December 2024). The $\le 30$-minute cadence of the TESS data, most notably in Sector 33 around the peak and caustic approach, provides powerful constraints on the binary geometry.

Owing to its survey strategy and sky coverage, TESS is not well suited for detecting classical microlensing events, and such events falling within the TESS footprint are expected to be rare \citep{SajadianTess, Yang_2024}. To the best of our knowledge, no previous confirmed microlensing event with TESS photometry has been reported in the literature, such that this event represents the first published case in which TESS photometry is used as part of the light-curve analysis of a gravitational microlensing event. Although \citet{kunimoto2025searchingfreefloatingplanetstess} proposed a free-floating planet candidate based on a single brightening event in a TESS light curve, subsequent analysis by \citet{mroz_2024} demonstrated that this signal is more consistent with a stellar flare than with microlensing.
        
\subsection{NEOWISE observations}
The NEOWISE reactivation mission \citep{NEOWISE} was the extended phase of the Wide-field Infrared Survey Explorer \citep[WISE;][]{WISE2010Wright}, which provided time series of mid-infrared imaging of the sky in the W1 (3.4 $\mu \mathrm{m}$) and W2 (4.6 $\mu \mathrm{m}$) bands. Primarily aimed at identifying and characterising near-Earth objects, NEOWISE delivered valuable time-domain data, which, for microlensing studies, offers the opportunity to probe the mid-infrared properties of the source and lens. At mid-infrared wavelengths, contributions from cool or dusty objects may be more readily detected than at optical or UV wavelengths. NEOWISE observations of the source span from March 2014 until February 2024. The source was routinely observed with a six-month cadence such that 21 NEOWISE-observing periods are available, of which four periods are during the time where the source was magnified.

%--------------------------------------------------------------------
\section{Spectroscopic observations}

\subsection{Observing data}
In addition to intensive photometric monitoring, Gaia20fnr was also observed spectroscopically by several instruments. We initiated spectroscopic follow-up with the Spectrograph for the Rapid Acquisition of Transients (SPRAT, \citealt{Piascik2014}) mounted on the 2-m Liverpool Telescope (LT, \citealt{Steele2004}) in La Palma, Canary Islands, Spain. The low-resolution spectrum (R$\sim$350) was taken on 23.12.2020 (proposal ID: OL20B01, PI: P. Zielinski) in the wavelength range $400-800 \, \mathrm{nm}$. The standard reduction procedure was applied by using an automated pipeline developed by the LT Team. The Xe arc lamp was used to calibrate the spectrum in the wavelengths.

We also collected low-resolution spectrum using Gemini Multi-Object Spectrographs (GMOS, \citealt{Hook2004}) mounted on the 8-m Gemini North Telescope (Hawaii, USA). The spectrum was obtained using the long-slit mode with gratings R400 which provides resolving power of R$\sim$2000 and covers the wavelength range from 400 to 900 nm. The observations were done on 09.01.2021 (proposal ID: GN-2020B-Q-318, PI: I. Gezer). The spectrum was bias- and flat-field corrected as well as wavelength and flux calibrated by using IRAF\footnote{\url{https://iraf.noirlab.edu/}} tasks and scripts. 

The third low-resolution spectrum (R$\sim$500) was obtained with the FLOYDS spectrograph on the LCO 2-m telescope at Siding Spring Observatory, Australia \citep{Brown2013}. FLOYDS covers wavelengths from roughly 320 to 1000~nm. A 1.2$\arcsec$ slit was used for these observations. The spectrum was obtained on 07.02.2021 (proposal ID: KEY2020B-003, PI: E. Bachelet). All reductions were carried out using the standard LCO FLOYDS pipeline where a set of HgAr and Zn lamps are used for wavelength calibration. Due to strong fringing at the red end of the detector, we restricted our analysis to wavelengths shorter than 750 nm. 

\begin{figure}
\centering
\includegraphics[width=\columnwidth]{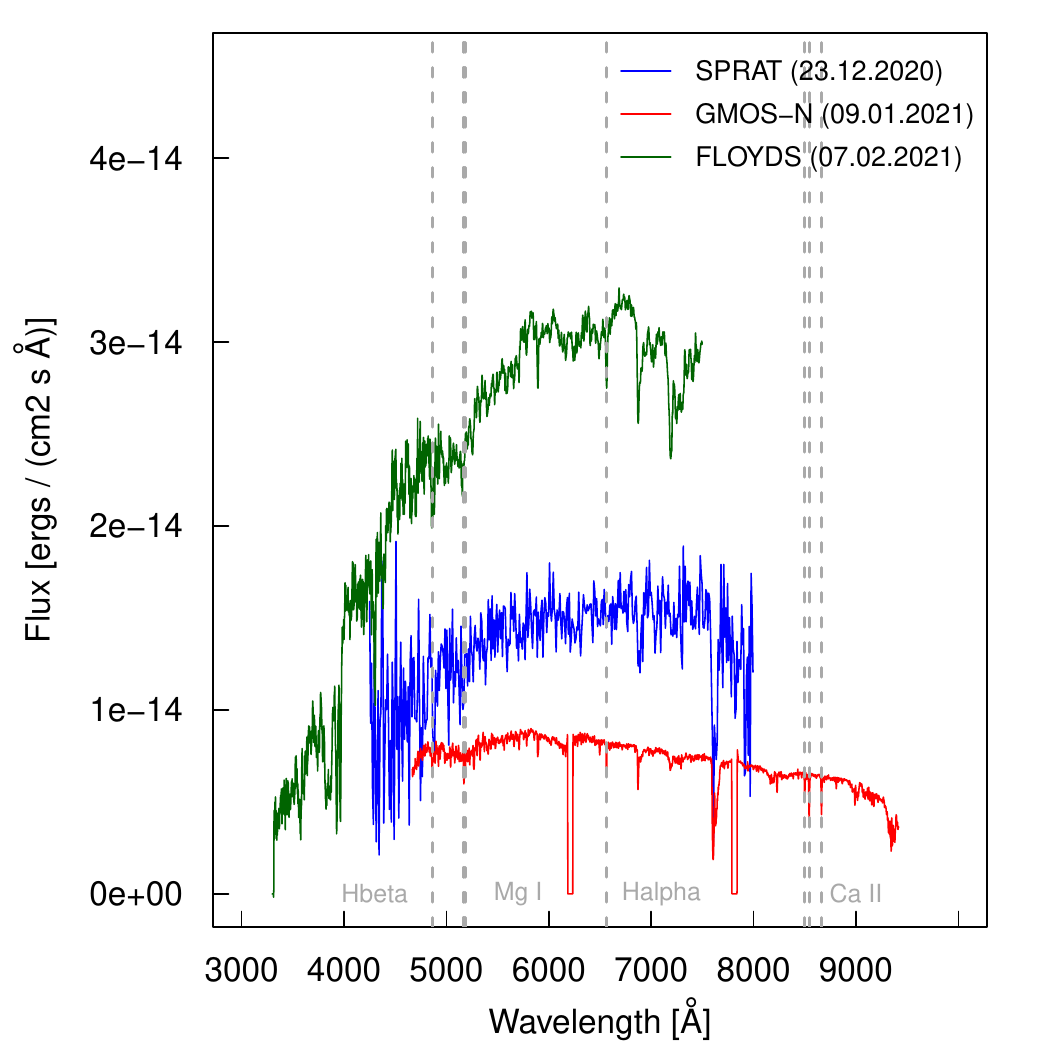}
\caption{Low-resolution spectra of the Gaia20fnr event observed by the SPRAT, GMOS-N and FLOYDS instruments. Vertical dashed lines present two Balmer lines, Mg I, and Ca II.}
\label{fig:low-res-spectra}
\end{figure}

All of these low-resolution observations, shown in Fig. \ref{fig:low-res-spectra}, revealed a spectrum characteristic of an ordinary late-type star, displaying strong metallic and Balmer absorption lines. No emission features were detected, indicating an absence of stellar activity or circumstellar material. Therefore, Gaia20fnr was identified as a candidate microlensing event \citep{Atel2021}.

For the high-dispersion spectra, we used the 10-m Southern African Large Telescope (SALT, \citealt{Buckley2006}). The observations were carried out on 06.02.2021 and 04.09.2021 (proposal ID: 2018-2-LSP-001, PI: D. Buckley). 
To obtain an {\'e}chelle spectrum covering both the blue and red channels (3700–8800 Å), we used the High-Resolution Spectrograph (HRS; \citealt{Crause2014}) in its medium-resolution configuration. This setup yielded resolving powers of $\mathrm{R}\sim15\,000$ and $\mathrm{R}\sim17\,000$ for the first and second spectra, respectively. The raw data were processed with the SALT MIDAS reduction pipeline \citep{Kniazev2016,Kniazev2017}, including bias subtraction, flat-field correction, and wavelength calibration using a ThAr arc lamp. The spectrum was then extracted into a one-dimensional format, order-merged, corrected for sky emission, and shifted to the heliocentric frame. We further refined the data by removing poor-quality parts, cosmic-ray hits and other artificial artefacts. Because the spectra exhibited low-quality at the blue end, regions with $\lambda < 4000 \AA$ (first spectrum) and $\lambda < 5480 \AA$ (second spectrum) were excluded. After these adjustments, the final spectra had an average signal-to-noise ratio of about $90$.

We also obtained observations with the X-Shooter instrument \citep{Vernet2011} on the ESO Very Large Telescope (VLT). X-Shooter is a multi-wavelength medium-resolution spectrograph equipped with three independent arms, enabling simultaneous coverage of the UVB (300–559 nm), VIS (559–1024 nm), and NIR (1024–2480 nm) spectral ranges. Because each arm uses its own cross-dispersing optics, detector shutter, and slit mask, the spectral resolution differs between them. Using slit widths of 1.0$\arcsec$, 0.7$\arcsec$, and 0.6$\arcsec$, we achieved resolution of $R\sim10\,000$ and SNR$\sim750$. X-Shooter observations were carried out on 02.10.2021 (ESO proposal ID: 108.22JZ.001, PI: L. Wyrzykowski). The data were processed with the EsoReflex pipeline\footnote{\url{https://www.eso.org/sci/software/esoreflex/}}. Wavelength calibration was performed with ThAr lamps for the UVB and VIS arms and a combination of Ar, Hg, Ne, and Xe lamps for the NIR. 

\begin{figure*}
\centering
\includegraphics[width=\columnwidth]{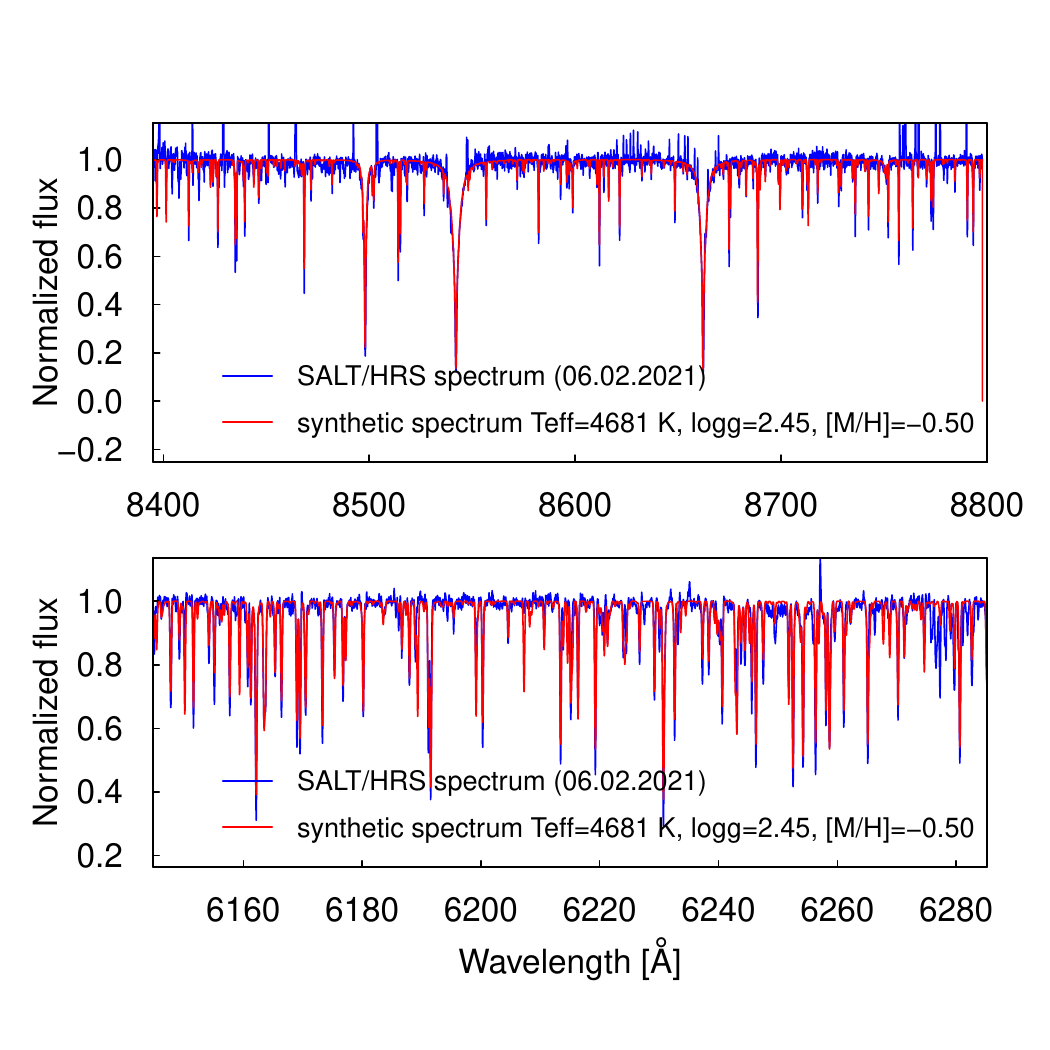}
\includegraphics[width=\columnwidth]{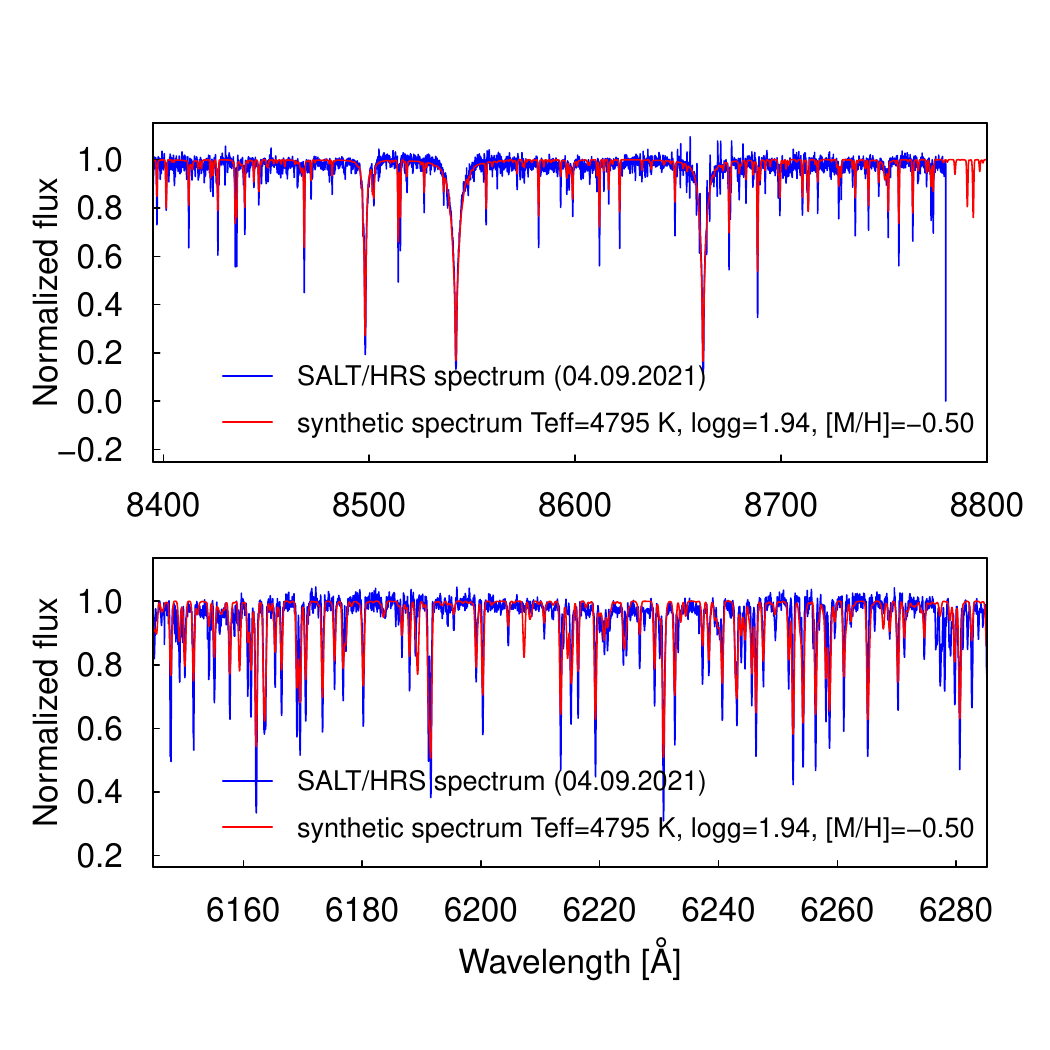}
\caption{Spectrum of the Gaia20fnr event (blue) observed by the SALT HRS spectrograph on the 06.02.2021 (left) and 04.09.2021 (right). Shown in red is the synthetic spectrum calculated for the best-fit atmospheric parameters.}
\label{fig:spectraSALT}
\end{figure*}

Parts of calibrated SALT/HRS and X-Shooter spectra are shown in Fig.~\ref{fig:spectraSALT} and Fig.~\ref{fig:spectraVLT}.

\subsection{Spectral analysis}
Based on three high-resolution datasets, we have analysed the absorption features in Gaia20fnr. The spectral resolution of these datasets is sufficient for determining the atmospheric parameters through the \textit{iSpec}\footnote{\url{https://www.blancocuaresma.com/s/iSpec}} package, which incorporates several known radiative-transfer tools \citep{BlancoCuaresma2014,BlancoCuaresma2019}. For this work, we employed the SPECTRUM code to derive the effective temperature $T_{\rm eff}$, surface gravity $\log g$, metallicity [M/H], and microturbulent velocity $v_{\rm t}$.

In order to generate synthetic spectra and compare them with the observations, we selected a list of atomic features including, e.g., $H_{\alpha}$, $H_{\beta}$, and lines of Ca, Mg, Na, Ti, and Fe, and used MARCS atmospheric models \citep{Gustafsson2008} together with solar abundances from \citet{Grevesse2007}. Since nearly all measurable lines with reliable laboratory data (wavelengths, excitation potentials, oscillator strengths, etc.) appear in the VIS arm of the X-Shooter spectrum, our analysis focuses exclusively on that wavelength range.

\begin{figure}
\centering
\includegraphics[width=\columnwidth]{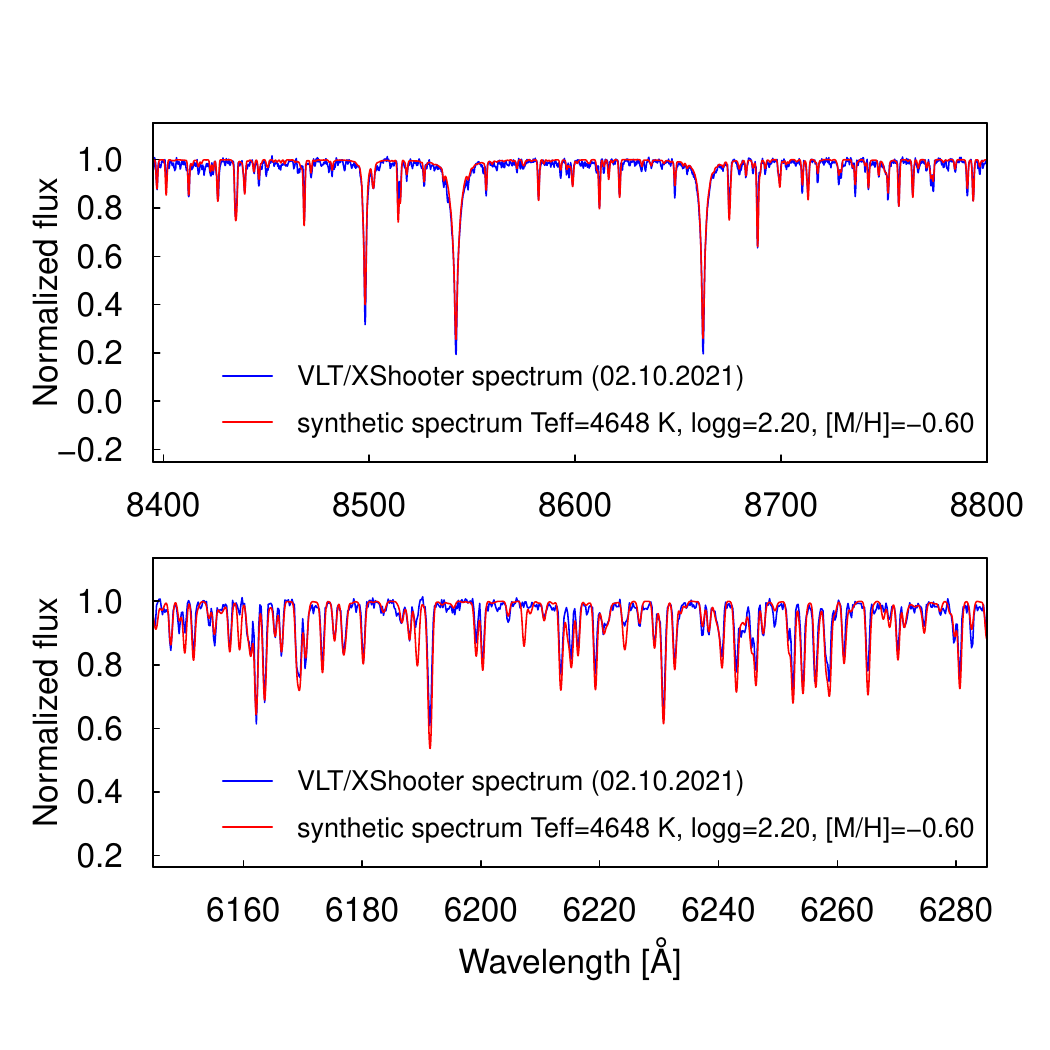}
\caption{Spectrum of the Gaia20fnr event (blue) observed by the X-Shooter instrument on the VLT on the 02.10.2021. Shown in red is the synthetic spectrum calculated for the best-fit atmospheric parameters.}
\label{fig:spectraVLT}
\end{figure}

The best-fitting synthetic spectra corresponding to the derived parameters are listed in Table~\ref{tab:stellar_parameters}. Figures~\ref{fig:spectraSALT} and~\ref{fig:spectraVLT} present the SALT/HRS and X-Shooter data and their synthetic counterparts around the Ca~II triplet, and the 6140–6280~$\AA$ region containing many Fe lines. Across all observations, the derived parameters are consistent within uncertainties. Based on them, we classified the source star of the Gaia20fnr microlensing event as a metal-deficient K2-type giant. Furthermore, none of the high-resolution spectra show any absorption features of a secondary component.

%\begin{table*}[h!]
%\centering
%\begin{tabular}{lcccccc}
%\hline
% \\[0.1ex]
%\textbf{Parameter} & \multicolumn{5}{c}{\textbf{Synthetic spectrum fitting}} & \textbf{Template matching} \\[1ex] 
% & 1st SALT & 2nd SALT & Gemini & 3rd SALT & VLT/4635
% 35
% &  \\[0.2ex] 
% & (23.12.2020) & (06.02.2021) & (09.01.2021) & (04.09.2021) & (02.10.2021) & \\
%\hline
% \\[0.2ex]
%$T_{\rm eff}$ [K] & - & $4681 \pm X$ & - & $4795 \pm X$ & $4648 \pm %X$ & - \\
%$\log g$ (cgs) & - & $2.45 \pm X$ & - & $1.94 \pm X$ & $2.20 \pm X$ %& - \\
%{[M/H]} [dex] & - & $-0.50 \pm X$ & - & $-0.50 \pm X$ & $-0.60 \pm %X$ & - \\
%$A_V$ [mag] & - & - & - & - & - & - \\
%\hline
%\\[1ex]
%\end{tabular}
%\caption{Stellar parameters derived from different methods and observations.}
%\label{tab:stellar_parameters}
%\end{table*}

\begin{table*}[h!]
\centering
\begin{tabular}{lccc}
\hline
% \\[0.1ex]
\textbf{Parameter} & \multicolumn{3}{c}{\textbf{Synthetic spectrum fitting}} \\[1ex] 
 & 1st SALT/HRS & 2nd SALT/HRS & VLT/X-Shooter  \\
 & (06.02.2021) & (04.09.2021) & (02.10.2021)  \\[1ex] 
\hline
% \\[0.2ex]
$T_{\rm eff}$ [K] & $4681 \pm 265$ & $4795 \pm 79$ & $4648 \pm 353$ \\
$\log g$ [cgs] & $2.45 \pm 0.48$ & $1.94 \pm 0.16$ & $2.20 \pm 0.40$ \\
{[M/H]} [dex] & $-0.50 \pm 0.30$ & $-0.50 \pm 0.10$ & $-0.60 \pm 0.28$ \\
$v_{\rm t}$ [$km\,s^{-1}$] & $1.25 \pm 0.59$ & $1.96 \pm 0.21$ & $2.49 \pm 1.08$ \\
\hline
\\[1ex]
\end{tabular}
\caption{Stellar parameters derived from synthetic spectrum fitting of high-resolution spectroscopic observations.}
\label{tab:stellar_parameters}
\end{table*}

% \begin{table}[h!]
% \centering
% \begin{tabular}{lccc}
% \hline
% % \\[0.1ex]
% \textbf{Parameter} & \multicolumn{3}{c}{\textbf{Synthetic spectrum fitting}} \\[1ex] 
%  & 1st SALT/HRS & 2nd SALT/HRS & VLT/X-Shooter  \\
%  & (06.02.2021) & (04.09.2021) & (02.10.2021)  \\[1ex] 
% \hline
% % \\[0.2ex]
% $T_{\rm eff}$ [K] & $4681 \pm 265$ & $4795 \pm 79$ & $4648 \pm 353$ \\
% $\log g$ (cgs) & $2.45 \pm 0.48$ & $1.94 \pm 0.16$ & $2.20 \pm 0.40$ \\
% {[M/H]} [dex] & $-0.50 \pm 0.30$ & $-0.50 \pm 0.10$ & $-0.60 \pm 0.28$ \\
% $v_{\rm t}$ [km/s] & $1.25 \pm 0.59$ & $1.96 \pm 0.21$ & $2.49 \pm 1.08$ \\
% \hline
% \\[1ex]
% \end{tabular}
% \caption{Stellar parameters derived from synthetic spectrum fitting of high-resolution spectroscopic observations.}
% \label{tab:stellar_parameters}
% \end{table}

%-----------------------------------------------------------------
\section{Microlensing model}
\subsection{Data preparation}
Due to the complexity of the microlensing event, the data used for modelling are restricted to the data shown in Fig.~\ref{fig:lcFit}. By selecting data from instruments that observed for substantial parts of the light curve, specifically regions during the magnification as well as at baseline, the efficiency of the fitting procedure dramatically increased. We emphasise that unmagnified data points do not impose constraints on the microlensing model parameters because the microlensing solution is determined through a non-linear fit.

\emph{Gaia} data (G-band) and ASAS-SN data (g-band) cover the entire magnification region as well as a substantial time during the baseline before and after the microlensing event, starting from before 2015 and 2018, respectively. LCO data in g- and i-bands as well as ROAD Observatory data in i-, B-, V-, and I-bands are well sampled during the source magnification periods and after the microlensing event returned to baseline.

For the modelling, the high cadence and high precision LCO datasets were vital in achieving model convergence. After removing instrumental signatures with the BANZAI pipeline \citep{BASNZAIMcCully}, time-series photometry is derived with the pyDANDIA software package \citep{Bramich2008, Bramich2013} and extracted to the reference frames \citep{street2024romereathreeyeartricolortimeseries}. Conventionally, the Sinistro instruments from different LCO observing sites can then be modelled as one single instrument; however due to the extensive data from the Cerro Tololo Inter-American Observatory, the South African Astronomical Observatory, and the Siding Spring Observatory, we choose to model them as individual telescopes/instruments with independent source-blend flux ratios. 
% Additionally, the high cadence and resolution of TESS photometry were key in tight orbital motion parameter constraints. The data around the first peak help to determine the precise model parameters.

To speed up the initial modelling processes, data were binned into 24-hour segments. Considering that there are no caustic crossings in this event, the binning was conservative and could have been with larger bins to speed up the initial modelling process. For the final models, data were unbinned and each instrument flux was modelled individually. The error bars of each dataset were rescaled according to common microlensing practice \citep{Skowron2011, Yee2012, Miyake2012}, such that 
\begin{equation}
\sigma_i = k \sqrt{\sigma_{i,\mathrm{orig}}^2 + e_{\mathrm{min}}^2} \, ,
\end{equation}
where $\sigma_{i,orig}$ is the original data error, and $e_{\mathrm{min}}^2$ and $k$ are such that $\chi^2 / \mathrm{dof} \approx 1$ and the cumulative sum of $\chi^2$ is approximately linear as a function of source magnification. The coefficients $e_{\mathrm{min}}^2$ and $k$ are determined independently for each dataset and the data were cut such that data with an error greater than three times the median were excluded. 

\subsection{Binary lens model}
We adopt the standard gravitational microlensing formalism, in which a foreground lens deflects the light of a background source and produces a time-varying magnification as the projected relative lens–source position changes. The characteristic angular scale of the problem is the angular Einstein radius,
\begin{equation}
\theta_{\mathrm{E}} = \sqrt{\frac{4\, G\, M_{\mathrm{L}}}{c^2}\,\frac{D_{\mathrm{S}} - D_{\mathrm{L}}}{D_{\mathrm{S}} \, D_{\mathrm{L}}}} \, ,
\label{equation:thetaE}
\end{equation}
where $M_\mathrm{L}$ is the total lens mass and $D_{\mathrm{L}}$ and $D_{\mathrm{S}}$ are the lens and source distances, respectively.

As a reference, we begin with the Paczy\'nski point-source point-lens (PSPL) model \citep{PaczynskiMicrolensing86}, defined by the parameters $\{t_0,\,u_0,\,t_{\mathrm{E}}\}$, where $t_0$ is the epoch of closest approach, $u_0$ is the minimum impact parameter in units of $\theta_{\mathrm{E}}$, and $t_{\mathrm{E}}$ is the Einstein timescale. The PSPL model can be extended to include higher-order effects such as the angular source radius in Einstein units ($\rho = \theta_\star/\theta_{\mathrm{E}}$) and the microlens parallax ($\pi_{\mathrm{E}} = \sqrt{\pi_{\mathrm{E},\mathrm{N}}^2 + \pi_{\mathrm{E},\mathrm{E}}^2}$). It provides a useful baseline to approximate $t_0$ but cannot reproduce the pronounced anomalies associated with the multi-peaked structure observed in the light curve.

The PSPL model is extended to the point-source binary-lens (PSBL) model following the standard parametrisation used throughout the microlensing literature \citep{Gaudi2012}. The binary lens extends the model by: (i) the projected separation, $s$, between the two lens components in units of $\theta_{\mathrm{E}}$; (ii) the mass ratio, $q = M_2/M_1$, of the binary lens; (iii) the angle, $\alpha$, between the source trajectory and the binary-lens axis.

The topology of the caustic curves of a binary lens depends primarily on $(s,q)$, and the close ($s\ll1$), resonant ($s\simeq1$), and wide ($s\gg1$) regimes are all in principle allowed \citep{ErdlSchneider, dominik1999binarygravitationallensextreme}. Viable solutions commonly appear in close--wide pairs owing to the $s\leftrightarrow1/s$ degeneracy.

Because the magnification $A(t)$ of a source due to gravitational microlensing is achromatic, the model flux in each dataset $i$ that describes a telescope and filter is written as
\begin{equation}
F_i(t) = A(t) \, F_{{\rm s},i} + F_{{\rm b},i} \, ,
\end{equation}
where $F_{{\rm s},i}$ and $F_{{\rm b},i}$ are the dataset-dependent source and blend flux, respectively. The geometric microlensing parameters are shared across all datasets, while the flux parameters are independent. The blend flux parameters contain contributions from the lens, an unrelated object in the line of sight, or both. This can provide constraints on the lens brightness and colour, which can reveal additional information about the nature of the lens. Given the long timescale of the event, magnified at $A > 1.05$ for more than 500 days, and its multi-platform coverage, higher-order effects are also considered.

\subsection{Higher-order effects}
\subsubsection{Annual and satellite microlensing parallax}

Annual parallax arises from Earth's orbital motion and is represented in the geocentric frame by the vector $\boldsymbol{\pi}_{\mathrm{E}} = (\pi_{\mathrm{E,N}}, \pi_{\mathrm{E,E}})$ such that
\begin{equation}
\pi_{\mathrm{E}} = \frac{\pi_{\mathrm{rel}}}{\theta_{\mathrm{E}}} \qquad \mathrm{and} \qquad
\pi_{\mathrm{rel}} = \mathrm{AU}\left(\frac{1}{D_{\mathrm{L}}} - \frac{1}{D_{\mathrm{S}}}\right),
\label{equation:piEpirel}
\end{equation}
following \citet{GouldJerk}. The microlens parallax introduces long-term distortions in the light curve, especially in the wings.

Space-based microlensing parallax, first proposed in \citet{refsdaslSpaceParallax}, is obtained by observing the same microlensing event simultaneously from Earth and a satellite \citep{Gould1994}. It has since proven to be observable and able to measure the microlensing parallax vector, from which the lens mass and distance are determined \citep[e.g.][]{Dong_2007_SPACEPARALLAX, Udalski_2015_SPACEPARALLAX, Yee_2015_SPACEPARALLAX}. Space-parallax signatures are particularly evident in data from space telescopes like Spitzer \citep{SpitzerPaper}, whose large Earth–spacecraft separation yields a strong parallax baseline. Nonetheless, they have also been measured in \emph{Gaia} microlensing events \citep[e.g.][]{Wyrzykowski16aye, Kruszynska2022Gaia18cbf}.

As the Gaia20fnr microlensing event was observed by four space telescopes, the ephemeris of each space telescope is considered in the light curve modelling. As the NEOWISE and Swift space telescopes are in low-Earth orbit this effect is not observable for this event with these telescopes. The TESS space telescope is in a highly‐elliptical, high‐altitude orbit, such that space-parallax is in theory observable; however it is not observed for this event. The \emph{Gaia} spacecraft, observing at Lagrange-Point L2, is most sensitive to the satellite parallax, and shows a space-parallax signal. This space-parallax is only observable at the closest approach of the source and the lens and it has an insignificant effect on the light curve and model parameters.

\begin{figure*}[t]
\centering
\includegraphics[width=\textwidth]{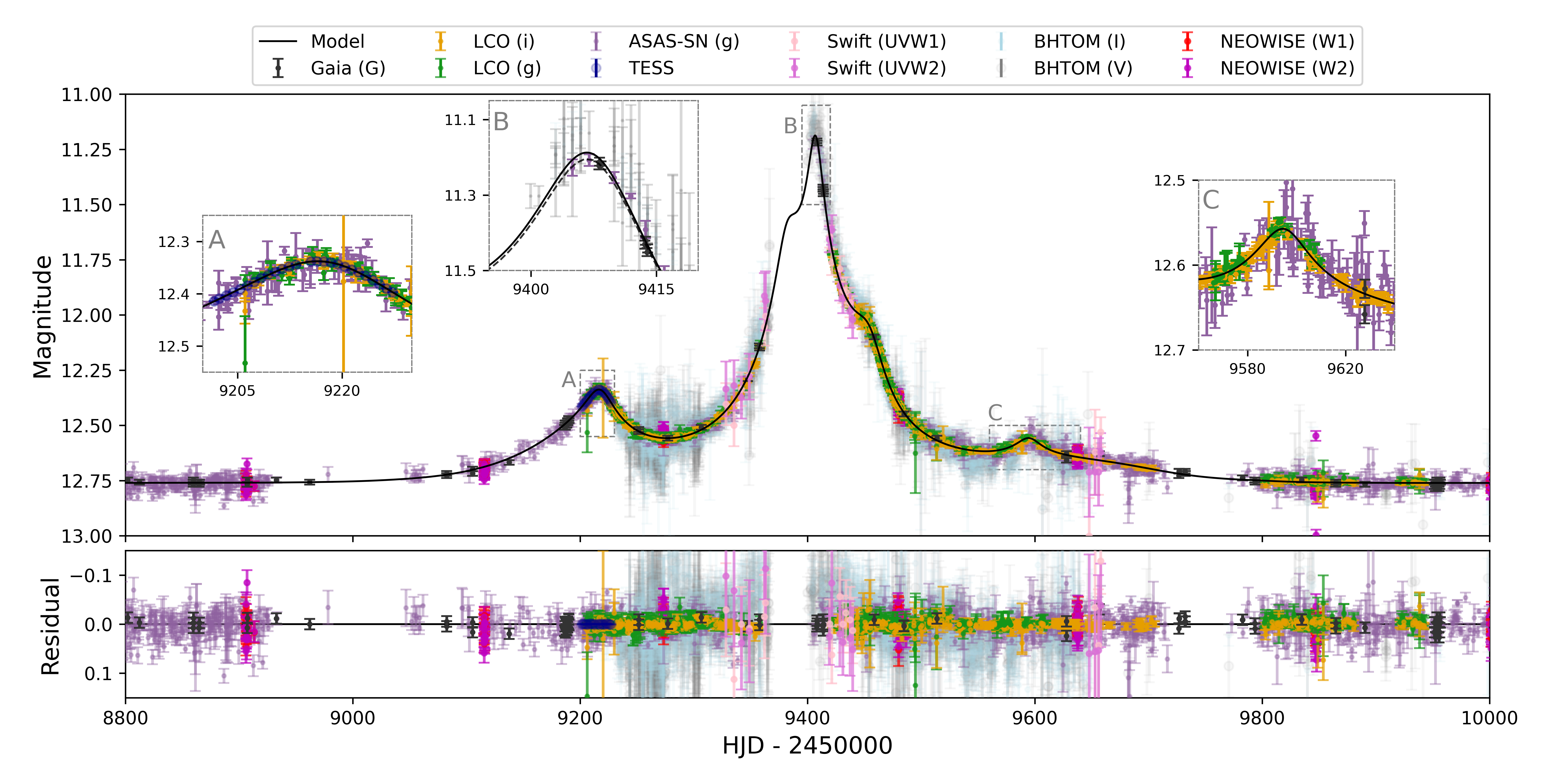}
\caption{Light curve of the microlensing event Gaia20fnr. Only the data used in the modelling process are shown and all measurements are transformed into the i-band magnitude scale. Data from different observatories, telescopes, and instruments are plotted together for the LCO, ASAS-SN, and BHTOM data. TESS data are binned into 24-hour bins for visibility reasons. In Panel B, the space-parallax effect for \emph{Gaia} is visible in the dashed line. In Panel C, the BHTOM data is left out for visibility reasons.}
\label{fig:lcFit}
\end{figure*}

\subsubsection{(no)-Finite source effects}
Finite source effects become detectable when the magnification gradient across the source star is large enough to be resolved in the light curve. This typically occurs during caustic crossings, where the magnification varies strongly over small angular scales \citep{YooFiniteSource, ZubLimbDarkening}. Although extracting these signals is more difficult in non-caustic-crossing binary events, finite-source signatures may also be recovered because the magnification pattern expands around cusp regions \citep{dominik1999binarygravitationallensextreme, GaudiGould}.

When finite-source effects are measurable, the combination of limb-darkened light-curve modelling and an independently estimated angular source radius $\theta_\star$ yields a direct estimate of the angular Einstein radius, $\theta_{\mathrm{E}} = \frac{\theta_\star}{\rho}$.

In our finite-source binary-lens (FSBL) modelling, we include the finite angular size of the source star following the formalism of \citet{Albrow_1999} with linear limb darkening implemented as described by \citet{Bachelet2012}. For this event, the light curve does not show evidence of finite-source effects: models including $\rho$ fit the data no better than point-source models and $\rho$ fails to converge. In such cases, the non-detection of finite-source effects imposes an upper limit $\rho \le \rho_{\mathrm{max}}$, and consequently imposes a lower bound on the lens mass 
\begin{equation}
M_{\mathrm{L}} \ge \frac{1}{\kappa \, \pi_{\mathrm{E}}} \, \left( \frac{\theta_{\mathrm{\star}}}{\rho_{\mathrm{max}}} \right) \, ,
\end{equation}
where $\kappa \equiv \frac{4G}{c^{2}\,{\rm AU}} \simeq 8.144~{\rm mas}\,M_{\odot}^{-1}$.
Since $\rho$ does not converge in finite-source fits, we adopt the 95th-percentile value as $\rho_{\mathrm{max}}$. For this event, the angular source radius is $\theta_\star = R_{\mathrm{S}} / D_{\mathrm{S}} \approx 12 [R_\odot] \, / \, 3.1 [\mathrm{kpc}] \approx 18 [\mu\mathrm{as}] $, which implies a lower limit on the lens mass of $M_{\mathrm{L}} \ge 0.12 \, \mathrm{M_\odot}$. This limit does not yield any additional useful constraint beyond those obtained from other methods.
% Size = 12 Rsun (TESS) at 3.1kpc, thus ($\theta_{\star} \sim 18 \, \mathrm{\mu as}$)

\subsubsection{Keplerian orbital motion}
Binary-lens microlensing events may deviate from the static-lens approximation if the projected separation or orientation of the lens components changes appreciably during the event. For sufficiently long timescales, orbital motion must be taken into account to avoid systematic biases in the inferred lens parameters \citep{dominik1999binarygravitationallensextreme, Albrow_2000}. As a first step, we include linear orbital motion by allowing the projected separation $s$ and the position angle $\alpha$ of the binary axis to vary with constant time derivatives, $ds/dt$ and $d\alpha/dt$. This formalism provides a first-order description of orbital effects when only a small fraction of the orbit is covered.

For Gaia20fnr, the event duration represents a non-negligible fraction of the expected orbital period of the binary lens. In this regime, a full Keplerian treatment is required. We therefore adopt the parameterisation by \citet{Skowron2011}, which was adapted by \citet{BozzaVBMicrolensing}, in which the lens system is described by its three-dimensional position and velocity at a chosen reference epoch in the lens centre-of-mass frame. The subsequent evolution of the system is obtained by integrating the Newtonian two-body problem, ensuring that all trial solutions correspond to dynamically consistent orbits.

The six-dimensional phase-space vector at the reference epoch can be transformed into the standard set of orbital elements of the binary. These are the semi-major axis $a$, orbital period $P$, eccentricity $e$, inclination $i$, longitude of the ascending node $\Omega$, argument of periapsis $\omega$, and time of periastron passage $t_{\mathrm{peri}}$. These elements uniquely define the three-dimensional orbit and, in particular, the projected positions of the two components on the sky as a function of time. In earlier binary microlensing events with measurable orbital motion, a Keplerian model typically provided only a modest improvement over the linear approximation \citep{Skowron2011, Shin_2012}, because the observed light curves covered only a limited portion of the orbit. In contrast, for Gaia20fnr the similarity between the orbital period and the event timescale means that the Keplerian description yields a significantly better fit than both the static and linear-orbit models and is therefore required for a physically reliable interpretation of the event \citep{Penny2011}.

We find that an additional degenerate configuration exists in our microlensing model, which differs only by a simultaneous sign reversal of \( r_\mathrm{s} \) and \( \gamma_{\mathrm{radial}} \), i.e. \((r_\mathrm{s}, \gamma_{\mathrm{radial}}) \rightarrow (- \, r_\mathrm{s}, - \, \gamma_{\mathrm{radial}})\). This alternative solution yields the same physical parameters, except that \( \Omega \) and \( \omega \) transform as \( \Omega \rightarrow \pi - \Omega \) and \( \omega \rightarrow \omega - \pi \), respectively, and the sign of the radial-velocity is reversed.

\subsection{Modelling methods and degeneracies}
Higher-order effects in the PSBL model are explored both individually and in combination to assess potential degeneracies inherent to photometric microlensing data. The modelling is carried out with the \texttt{pyLIMA} software package \citep{pyLIMApaper}, which provides a flexible, modular framework for microlensing analysis and offers interfaces to fast binary-lens magnification solvers. In particular, \texttt{pyLIMA} employs the \texttt{VBMicrolensing} backend \citep{BozzaVBMicrolensing, Bozza_2025}, which implements an advanced contour-integration algorithm for accurate computation of finite-source binary-lens magnifications. \texttt{pyLIMA} has been successfully used to model a variety of microlensing events \citep[e.g.][]{Bachelet20bof, Bachelet2018OGLE0417}.

The binary-lens parameter space is high-dimensional and can contain multiple local minima associated with known degeneracies. To explore these minima robustly, we initially adopt a strategy for a global search using a Differential Evolution (DE) algorithm. DE is a population-based stochastic optimization method \citep{storn1997differential} well suited for multi-modal problems. We initialise the DE population across wide ranges in all fit parameters. The \texttt{pyLIMA} algorithms evaluate the log-likelihood across all datasets, solving analytically for $(F_{{\rm s},i},F_{{\rm b},i})$ for speed. After convergence, we identify the families of solutions (e.g. close vs.\ wide, positive vs.\ negative $u_0$) and retain all models within a small $\Delta\chi^2$ of the best minimum to seed the Markov Chain Monte Carlo (MCMC) stage. We then explore the posterior distribution around each DE minimum with MCMC as implemented in \texttt{pyLIMA} via \emph{emcee} \citep{emceePaper}.

\subsection{Best-model results}

\begin{figure}
\centering
\includegraphics[width=\columnwidth]{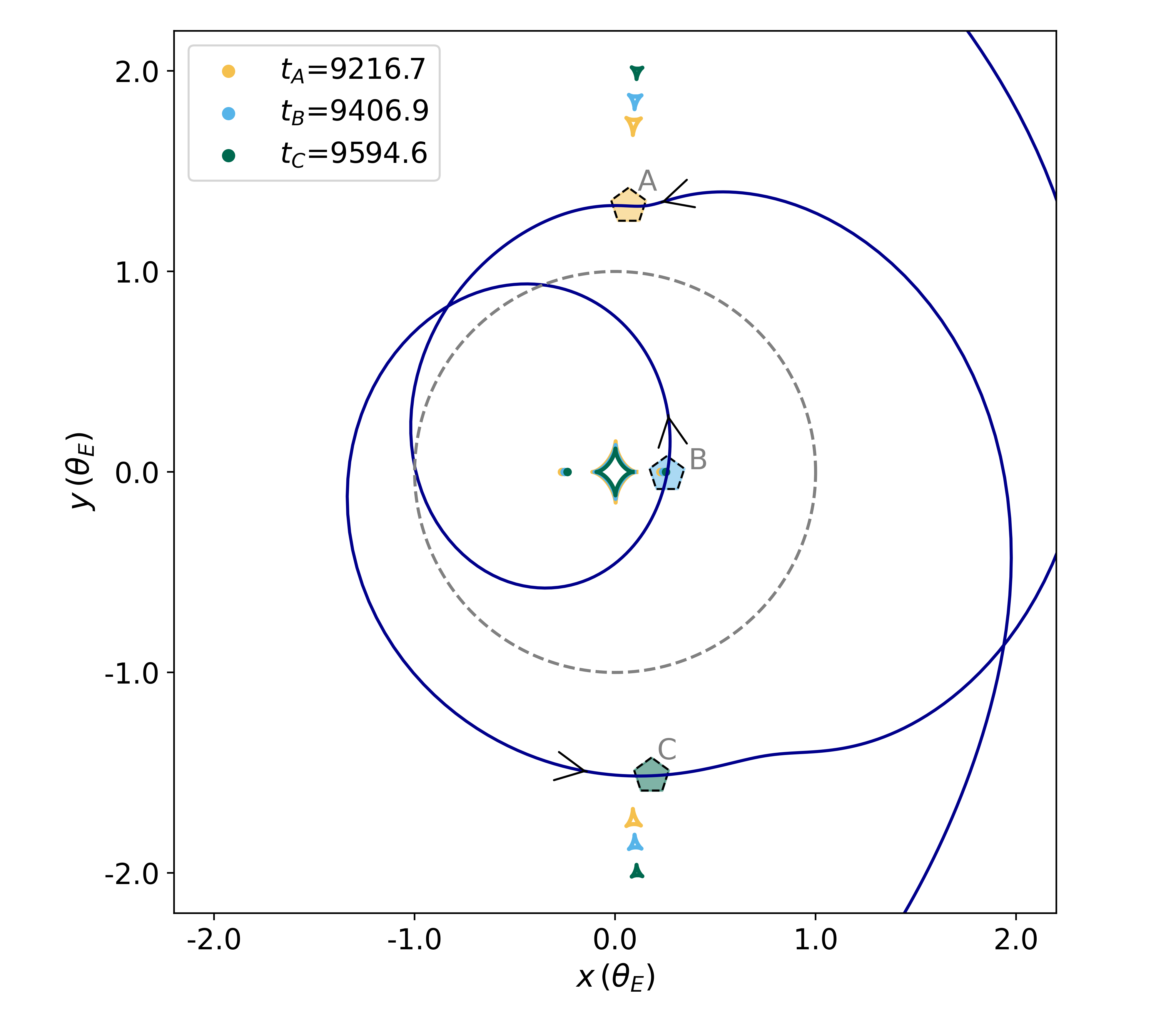}
\caption{Source trajectory and caustic geometry of the best-fit model shown for three timestamps. The dark blue line represents the source motion in the lens plane, strongly characterised by the parallax and orbital motion signals. The filled circles at $y(\theta_\mathrm{E}) = 0.0$ represent the lens positions. The outlined shapes mark the caustic structures. The different structures arise as a consequence of the orbital motion of the lens. The shaded pentagon markers represent the positions of the source in the lens plane at the three timestamps.}
\label{fig:causticGeometry}
\end{figure}

The best-fit result for the close ($s < 1$) binary-lens model is shown in Table~\ref{table:bestModel}. The MCMC chains only converged for the PSBL model that considered parallax and full Keplerian orbital motion. The source and blend magnitudes for the best model, shown in Table~\ref{table:magnitudesBM}, predict a flux ratio
% \begin{equation}
% \left( \frac{f_{\mathrm{source}}}{f_{\mathrm{source}} + f_{\mathrm{blend}}} \right) \approx 0.98 \,\,\,\, , \,\,\,\, \left( \frac{f_{\mathrm{blend}}}{f_{\mathrm{source}}} \right) \approx 0.02
% \end{equation}
\[
\left( \frac{f_{\mathrm{source}}}{f_{\mathrm{source}} + f_{\mathrm{blend}}} \right) \simeq 0.98 \,\,\,\,;
\,\,\,\,
\left( \frac{f_{\mathrm{blend}}}{f_{\mathrm{source}}} \right) \simeq 0.02
\]
in optical bands. NEOWISE observations determine a flux ratio of $0.985$ and $0.970$ in the W1 and W2 bands, respectively. Swift optical bands are in agreement with other optical observations like \emph{Gaia} and LCO; however, the UV bands have a flux ratio $\sim 1.0$, indicating that there is no significant UV light observed from the blending object. The light curve and best-fitting model, showing the unique, non-caustic-crossing structure, are presented in Fig.~\ref{fig:lcFit}. The geometry of the event and the motion of the source projected into the lens plane is shown in Fig.~\ref{fig:causticGeometry}.

The modelling approach outlined above provided no solution with a wide ($s > 1$) geometry. Similarly, binary source models with xallarap cannot reproduce the observed light curve.

\begin{table}[ht]
\caption{Microlensing parameters and errors corresponding to the 16, 50, and 84 percentiles of the MCMC chains of the best model result.}
\centering
\renewcommand{\arraystretch}{1.3}
\begin{tabular}{lr}
\hline
\textbf{Parameter} & \textbf{PSBL Model} \\
\hline
$t_0$ [JD-2450000] & $9403.881_{-0.098}^{+0.095}$ \\
$u_0$ & $-0.255_{-0.002}^{+0.002}$ \\
$t_\mathrm{E}$ [days] & $69.418_{-0.401}^{+0.403}$ \\
$s$ & $0.492_{-0.003}^{+0.003}$ \\
$q$ & $1.134_{-0.017}^{+0.016}$ \\
$\alpha$ & $4.496_{-0.005}^{+0.005}$ \\
$\pi_{\mathrm{E,N}}$ & $-0.185_{-0.001}^{+0.001}$ \\
$\pi_{\mathrm{E,E}}$ & $0.395_{-0.003}^{+0.003}$ \\
$\gamma_{\parallel}$ & $1.577_{-0.050}^{+0.047}$ \\
$\gamma_{\perp}$ & $-2.640_{-0.061}^{+0.060}$ \\
$\gamma_{\mathrm{radial}}$ & $-4.704_{-0.020}^{+0.022}$ \\
$r_\mathrm{s}$ & $0.378_{-0.016}^{+0.015}$ \\
$a_\mathrm{s}$ & $0.982_{-0.002}^{+0.002}$ \\
\hline    
$\chi^2 / \mathrm{dof} $ & 2246.9 / 2247.0 \\
\hline
\end{tabular}
\label{table:bestModel}
\end{table}

\begin{table}[ht]
\caption{Magnitude parameters and errors corresponding to the 16, 50, and 84 percentiles of the MCMC chains of the best model result. In brackets, the $3\sigma$ errors are included ((0.135, 99.865)-percentiles).}
\centering
\renewcommand{\arraystretch}{1.5}
\begin{tabular}{lrr}
\hline
\textbf{Filter} & \textbf{Source} & \textbf{Blend} \\
\hline
$G$ & $13.18_{-0.01}^{+0.01} \left(_{-0.03}^{+0.03}\right) $ & $17.63_{-0.50}^{+1.00} \left(_{-1.16}^{+9.77}\right)$ \\
$g$ & $13.90_{-0.01}^{+0.01} \left(_{-0.03}^{+0.03}\right)$ & $17.78_{-0.31}^{+0.43} \left(_{-0.74}^{+2.21}\right)$ \\
$i$ & $12.80_{-0.01}^{+0.01} \left(_{-0.03}^{+0.03}\right)$ & $16.63_{-0.29}^{+0.40} \left(_{-0.77}^{+2.83}\right)$ \\
\hline
\end{tabular}
\label{table:magnitudesBM}
\end{table}

%-----------------------------------------------------------------
\section{Physical Parameters}    
\subsection{\emph{Gaia} DR3 Parameters}
The alert of the microlensing event occurred on 2 December 2020 and our best model predicts that a magnification $A > 1.05$ did not occur before September 2020. The lens-source separation did not cross the unit Einstein radius ($u = 1$) until December 2020. \emph{Gaia} DR3 is based on data obtained until mid 2017, with the last observations for ID 2990431491637998848 made on 2 May 2017. Although the final observations used for the five-parameter solutions published in \emph{Gaia} DR3 were more than three years before the start of the photometric microlensing event, one cannot rule out that the astrometric microlensing signal influenced the source parameters published in \emph{Gaia} DR3 \citep[e.g.][]{Rybicki2018, Kruszynska2022Gaia18cbf, Maskoliunas2024}. Additionally, the best model results with non-zero blend flux mean that \emph{Gaia} flux measurements come from the combined flux of the source and the blend. This could be another potential contaminant to \emph{Gaia} DR3 5-parameter solutions. Thus, even for microlensing events that begin well outside the \emph{Gaia} DR3 timeframe, it is essential to scrutinise the \emph{Gaia} DR3 source parameters before using them in further analysis.

\begin{figure}
    \centering
    \includegraphics[width=\columnwidth]{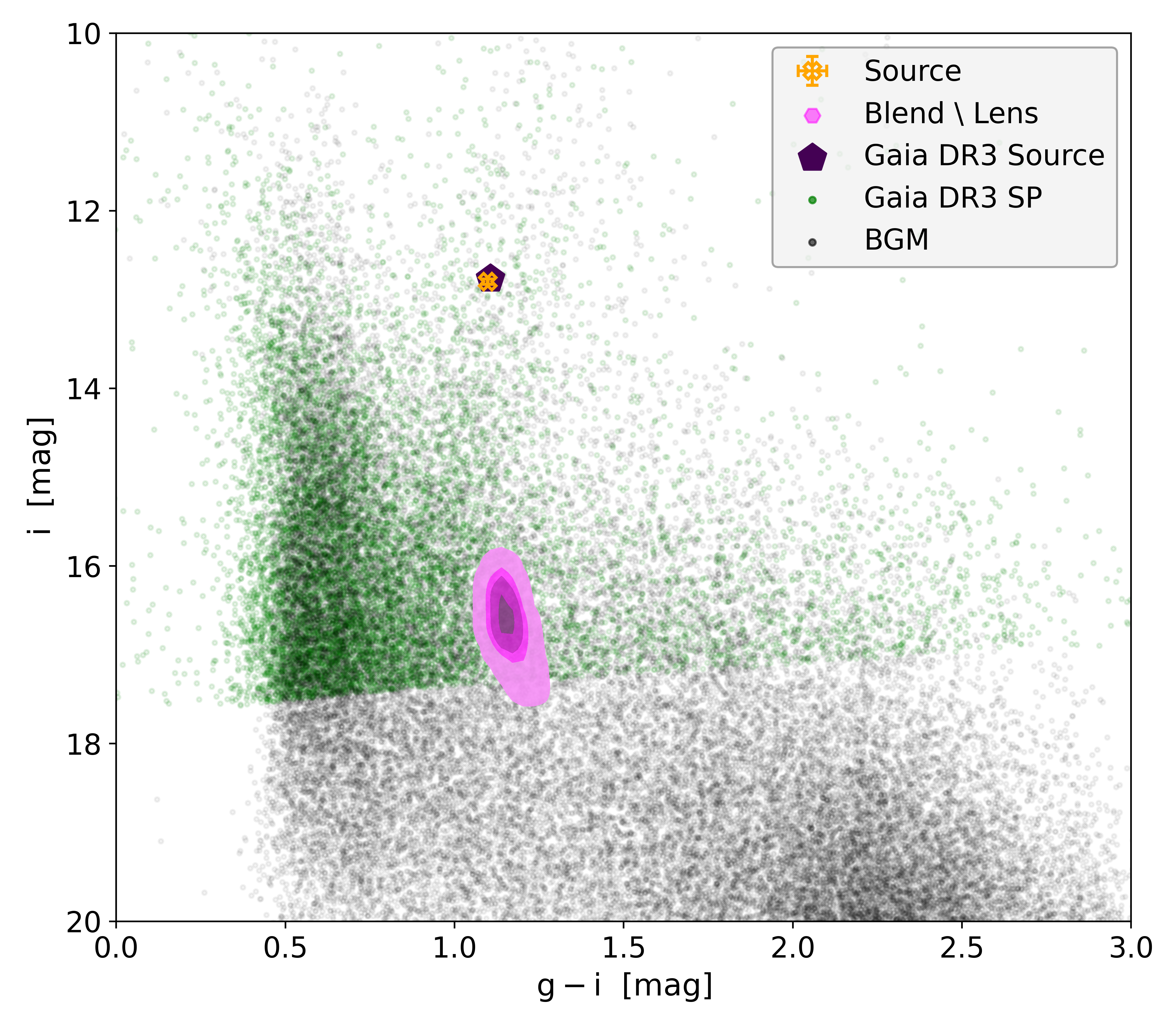}
    \caption{Colour Magnitude Diagram. Black dots are Besan\c{c}on Galaxy Model (BGM) neighbours within 1 degree of the source \citep{RobinBGM, CzekajBGM}. Green dots are the \emph{Gaia} DR3 Synthetic Photometry Catalogue neighbours within 1 degree of the source \citep{MontegriffoGaiaSPC}. The purple pentagon is the \emph{Gaia} DR3 Source of the Gaia20fnr GSA alert \citep{GaiaDR3}. The orange x-shaped marker shows the source position for the best model photometry. The pink contours indicate the blend position corresponding to the best-fitting photometric model, with contour levels enclosing the 20th, 40th, 60th, and 80th percentiles of the MCMC posterior.}
    \label{fig:cmdplot}
\end{figure}

The flux ratio determined in \emph{Gaia} G-band is such that the flux-weighted parallax, determined as the combined signal of source and blend parallax \citep{Rota21blx}, does not deviate significantly from the value published in \emph{Gaia} DR3. The Colour-Magnitude-Diagram (CMD), in Fig.~\ref{fig:cmdplot}, shows that the best-model source flux in SDSS-bands is in good agreement with the \emph{Gaia} DR3 Synthetic Photometry Catalogue (GSPC; \citet{MontegriffoGaiaSPC}) values of the source. In the CMD, the source lies in or near the RC region, consistent with the $12.16 \, \mathrm{R_\odot}$ radius inferred from TESS and with the K-type giant parameters reported in \emph{Gaia} DR3. A RUWE of 1.03 and a significant, positive, and non-zero parallax, $\varpi = 0.2875 \pm 0.0125 \, \mathrm{mas}$, imply a robust astrometric solution. We therefore assume that the values published in \emph{Gaia} DR3 are consistent with measurements of the source of the microlensing event.
    
\subsection{Source Distance, Extinction, and Type}
To determine the physical properties of the lensing event (Eq.~\ref{equation:thetaE}), a distance estimate for the source star ($D_S$) is essential. The source star distance published for the \emph{Gaia} DR3 source is determined from the parallax as $D_{\mathrm{s}} = 3099 \pm109 \,\mathrm{pc}$. The \citet{2021Bailer-Jones} catalogue distances, which were calculated based on \emph{Gaia} parallaxes and photometric measurements, are in agreement with this value. The geometric distance, based on the parallax and its uncertainties, gives the distance to the source star $3016_{-123}^{+130} \, \mathrm{pc}$. The photo-geometric distance is determined by the parallax, the colour, and the observed magnitude of the star, and gives a distance $3109_{-113}^{+97} \, \mathrm{pc}$. For the following calculations the source distance from the \citet{2021Bailer-Jones} catalogue, $3109_{-113}^{+97} \, \mathrm{pc}$, is used.

Given the significant extinction toward Galactic bulge microlensing fields, red clump (RC) giants in the CMD serve as standard candles for estimating the interstellar extinction along the line of sight \citep{YooFiniteSource, natafExtinction, BensbyExtinction}. In regions as isolated as the Gaia20fnr event, no statistically significant RC population, located along the red giant branch of the CMD, can be clearly separated from the main sequence. We therefore use the excess colour between the modelled source fluxes and the intrinsic colour estimated from the spectroscopic classification of the source to constrain the extinction in the line of sight of the microlensing event. The event lies at a relatively high Galactic latitude where the dust column density, and thus the extinction, is markedly lower than toward the bulge microlensing fields. Even so, the extinction is not negligible and must be included when determining the intrinsic source and blend colours and luminosities.

From \citet{Houdashelt2000}, a star with stellar parameters as derived from spectroscopic observations (Table~\ref{tab:stellar_parameters}) has the intrinsic colour $(V-I)_0 = 1.016 \pm 0.010 \,\mathrm{mag}$. The error accounts for the uncertainty in the spectroscopically derived parameters. Using the colour–colour relations from \citet{Jordi2006}, the intrinsic K2-type giant colour can be converted to an intrinsic SDSS colour of $(g-i)_0 = 0.974 \pm 0.010 \,\mathrm{mag}$. 
The source magnitudes in the g and i bands are constrained to within $± 0.01 \, \mathrm{mag}$ using data from the 1-m Sinistro telescopes of LCO at Siding Spring Observatory. From these measurements, the excess colour is estimated as $E(g-i) = (g-i)_{\mathrm{observed}} - (g-i)_{\mathrm{intrinsic}} = 0.125_{-0.015}^{+0.015}$. This corresponds to an extinction of $A_\mathrm{g} = 0.2570_{-0.019}^{+0.019}\,\mathrm{mag}$ and $A_\mathrm{i} = 0.1321_{-0.010}^{+0.010}\,\mathrm{mag}$. Applying these values to the LCO observations from Siding Spring yields corrected source magnitudes of $m_{\mathrm{g, LCO-SS}} = 13.64_{-0.02}^{+0.02} \,\mathrm{mag}$ and $m_{\mathrm{i, LCO-SS}} = 12.67_{-0.01}^{+0.01}\,\mathrm{mag}$.

In combination with the standard extinction coefficients for SDSS filters from a standard reddening law ($c_\mathrm{g} = 3.303$ and $c_\mathrm{i} = 1.698$) \citep{CardelliExtinction, SchlaflyFinkbeiner}, the relation ${E(g-i)}/{E(B-V)} = 1.605$ predicts the extinction of the source in JC bands. These approximate transformations from the SDSS to JC photometric system \citep{Jordi2006, SchlaflyFinkbeiner} predict a V-band extinction of $A_\mathrm{V} = 0.24 \pm 0.02 \,\mathrm{mag}$, which we can compare to other sources for an independent confirmation of our derived extinction parameters. % https://irsa.ipac.caltech.edu/cgi-bin/bgTools/nph-bgExec

The commonly used extinction maps of \citet{Marshall2006} do not cover regions located as far from the Galactic centre as is the case for this event. The extinction value in \citet{SchlaflyFinkbeiner}, derived from the colours of stars with spectra in the SDSS, are $A_\mathrm{V} = 0.224 \,\mathrm{mag}$. The dust maps of \citet{SchlegelFinkbeinerDouglas} predict $A_\mathrm{V} = 0.2605 \,\mathrm{mag}$. The Besan\c{c}on Galaxy Model (BGM) \citep{RobinBGM, CzekajBGM} is a population synthesis framework that simulates the stellar content and structural components of the Milky Way based on observational constraints for Galactic formation, evolution, and kinematics. It provides predictions of stellar distributions and observable properties along any line of sight and is widely used for interpreting survey data and modelling Galactic stellar populations in microlensing studies \citep[e.g.][]{Penny2013, Bachelet2022MOAExoplanet}. The BGM derives extinction data from a diffuse extinction model and predicts the extinction in the direction of the microlensing event as $A_{\mathrm{V, BGM}} = 0.260_{-0.024}^{+0.026}$ for the source distance. This is in agreement with the values derived above.
% PZ: add errorbars and compare Av with literature

Additionally, we use optical archival photometry data from the TESS catalogue \citep{TESS_2019} to verify the extinction and distance values for the K2 giant source star. The unmagnified source star was measured with $B = 14.442 \pm 0.023 \, \mathrm{mag}$., and $V = 13.356 \pm 0.057 \, \mathrm{mag}$. According to \citep{Straizys1992}, the intrinsic colour of the K2 giant should be $(B-V)_0 = 1.16 \, \mathrm{mag}$ with a typical absolute magnitude $M_\mathrm{V}=0.7 \pm 0.3 \, \mathrm{mag}$. For the standard reddening law in the diffuse ISM of the Milky Way, we assume $R_\mathrm{V} \sim 3.1$ \citep{CardelliExtinction} such that the estimated colour excess value is $E_{B-V} = 0.075 \pm 0.061$ and the extinction is $A_\mathrm{V} = 0.23 \pm 0.19\,\mathrm{mag}$. The spectroscopic distance calculated with the TESS photometry is $D_\mathrm{S} = 3.06 \pm 0.77 \, \mathrm{kpc}$, which is in agreement with the values published in \emph{Gaia} DR3.

We determined the kinematic properties of the source star by integrating its orbit over 5~Gyr within a theoretical Milky Way potential, using the Python-based galactic dynamics package \texttt{galpy} \citep{Bovy15}. For the integration, we adopted the gravitational potential \texttt{MWPotential2014}, which combines components for the bulge, disk, and halo. Proper motions, radial-velocity, and sky coordinates were taken from the \emph{Gaia} catalogue, and the source distance from this work. From these data, we derived a mean galactocentric distance of $R_{\mathrm{mean}} = 13.7\pm 0.37~\mathrm{kpc}$, a maximum vertical height from the Galactic plane of $|{\rm z_{\mathrm{max}}}| = 1.75\pm0.09~\mathrm{kpc}$, and an orbital eccentricity of $e = 0.25\pm0.01$. Our kinematic analysis indicates that the source star is most likely a thin-to-thick disk object.

To estimate the source age and mass, we employed \texttt{UniDAM}, which uses a Bayesian approach combined with \texttt{PARSEC} isochrones \citep{Bressan12}, stellar atmospheric parameters (derived from our spectroscopic analysis), and infrared magnitudes \citep{Mints17, Mints18}. The required \textit{J}, \textit{H}, and \textit{K} magnitudes from the 2MASS Survey and the $W1$ and $W2$ magnitudes from AllWISE were provided as input to \texttt{UniDAM}. The resulting age and mass of the source star are $4.7_{-1.3}^{+3.7}~\mathrm{Gyr}$ and $1.2\pm0.3~M_\odot$, respectively.

% astroARIADNE

\subsection{Physical parameters of the binary lens}
A direct lens mass estimate based on the microlensing model is given by 
\begin{equation}
M_\mathrm{L} = \frac{\theta_\mathrm{E}}{\kappa \, \pi_\mathrm{E}} \, .
\end{equation}
As the FSBL best-fit model does not constrain the finite source effects, a direct calculation of the lens mass and distance is not possible. Instead, if an estimate of the lens distance can be made, the lens mass can be calculated as
\begin{equation}
M_\mathrm{L} = \frac{\pi_{\mathrm{rel}}}{\kappa \, \pi_\mathrm{E}^2} .
\end{equation}
In cases where the finite source effects are not constrained, microlensing studies have often relied on Galactic models to constrain the parameters of the lens \citep[e.g.][]{Bachelet2022MOAExoplanet, Kruszynska2022Gaia18cbf, Howil2025Gaia18ajz}. For equal-mass binaries with  $q \sim 1$, as is the case for the microlensing event in this study, these Galactic models tend to underestimate the mass of the lens due to the underlying parameter relations being calibrated on single-star populations. 

DarkLensCode (DLC) \citep{Howil2025Gaia18ajz} estimates the probability distributions of lens mass, distance, and luminosity in microlensing events from microlensing parameters and Galactic population priors. To test its effectiveness in determining lens parameters for this event, we run the analysis for model parameters excluding the blend flux. This method estimates the lens mass as $0.98 \pm 0.38 \, \mathrm{M_\odot}$ in the 16th to 84th percentile range, which is in agreement with the values determined from our methods outlined below. At the $3\sigma$ level, DLC does not add additional constraints on the lens mass and distance beyond those imposed directly by the model parameters. It is therefore not considered further in this analysis.

\begin{figure}
    \centering
    \includegraphics[width=\columnwidth]{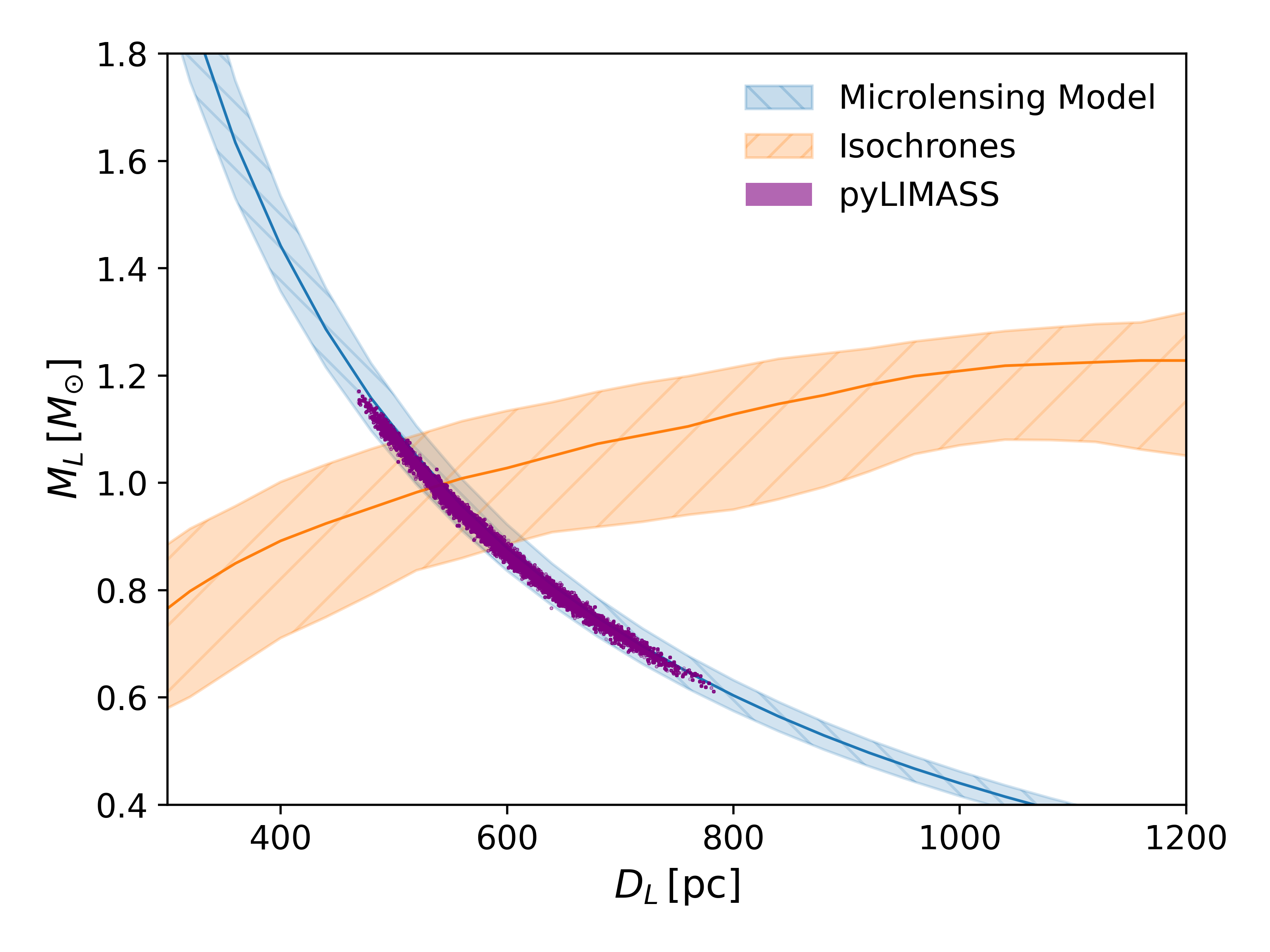}
    \caption{Lens distance vs mass plot with distributions for our three avenues of lens mass estimation. The  \emph{Microlensing Model}-strip shows values determined based on constraints from the physical microlensing model. The \emph{pyLIMASS}-distribution shows the posterior chains of the MCMC-anlysis with \emph{pyLIMASS} \cite{pylimassPaper}. The \emph{Isochrones}-strip shows the values determined for PARSEC isochrones \citep{PARSECpaper, PARSECpaper2}, assuming the blend light is coming from the lens.}
    \label{fig:lensMassPlot}
\end{figure}

We identify three avenues to constrain the lens mass. The lens mass as a function of the lens distance, constrained by these three methods, is shown in Fig.~\ref{fig:lensMassPlot}.

Firstly, the microlensing parameters in Equations~\ref{equation:thetaE} and \ref{equation:piEpirel}, specifically the microlensing parallax $\pi_\mathrm{E}$, predict the lens mass as a function of lens distance. Secondly, \texttt{pyLIMASS} \citep{pylimassPaper}, an algorithm that leverages information from the microlensing event with stellar isochrones in a Gaussian mixture approach, is implemented. For this event, \texttt{pyLIMASS} is able to set strong bounds on the mass-distance relation determined only by considering the constraints from the microlensing parameters. Lastly, stellar isochrones can be used to unravel the nature of the blend flux, which, assuming that the blend light is coming from the lens, can help identify the lens nature.

The blended light detected in a microlensing event could either come from the lensing objects, from neighbouring objects that are unresolved in the field of view but not magnified by the lens, or from a combination of both. Given the isolated nature of the Gaia20fnr microlensing event (see Fig.~\ref{fig:milkywayViewplot}), the most probable case is that the lens light causes the modelled blend flux. We therefore assume that the blended light comes solely from the binary lens object and will show that this is a self-consistent assumption. As the distance to the lens is not known, we do not assume that the source and lens are equally reddened, nor do we assume that the lens suffers from no extinction. The extinction of the BGM is modelled as a function of distance using a logistic growth function and is used to deredden the blend magnitude. The CMD (Fig.~\ref{fig:cmdplot}) illustrates that the (g-i) colour of the blend is well constrained despite significant uncertainties in the magnitude presented in Table~\ref{table:magnitudesBM}. It is important to note that the blend flux consists of the light from both binary objects. The distribution of blend/lens colour and magnitude is shown in the CMD (Fig.~\ref{fig:cmdplot}).

The dereddened photometric results of the blend can be leveraged with stellar isochrones from PARSEC \citep{PARSECpaper2, PARSECpaper}, and the mass ratio determined in the microlensing model to constrain the mass of the binary lensing object. We infer the component masses and distance of an unresolved binary microlens by combining the microlensing parallax constraint with the unresolved g/i-band fluxes interpreted through stellar isochrones at a fixed mass ratio \(q=M_2/M_1\). For a set of distances, defined in the line of sight of the observer to the source, each scenario is computed. The extinction is determined as a function of distance using the parametric fit of extinction values in the BGM. The apparent magnitude of the combined binary lenses is determined as
\begin{equation}
    M_\mathrm{g}^{\mathrm{obs}}(D_\mathrm{L}) = g_0(D_\mathrm{L}) - 5\log_{10}(D_\mathrm{L} \, / \, 10{\rm\,\mathrm{pc}})
\end{equation}
where $ g_0 = m_\mathrm{g} - A_\mathrm{g}(D_\mathrm{L})$. For a stellar isochrones grid of metallicity ($[\mathrm{Fe/H}] \in [-1.0, 0.5]$) and ages ($1 - 12$ Gyr), we build interpolators for single-star absolute magnitudes as a function of mass \(M_\mathrm{g}(\mathrm{Mass}), M_\mathrm{i}(\mathrm{Mass})\). The unresolved absolute magnitudes of the binary in band $X\in\{g,i\}$ are 
\begin{equation}
M_{X}(\mathrm{Mass}_{\mathrm{tot}}) = -2.5\log_{10}\!\Big(10^{-0.4M_X(\mathrm{Mass_1})}+10^{-0.4M_X(\mathrm{Mass}_2)}\Big) \, .
\end{equation}
In order to avoid confusion in the notation, the absolute magnitudes are marked as $M$ and the lens-component masses as $\mathrm{Mass}$.
We are able to determine the absolute magnitude of our binary lens in the g- and i-bands as a function of the distance to the binary lens. By interpolation, this estimates the total mass with a fixed mass ratio $q$, such that the individual masses of the binary lens can be determined.

\subsubsection{Summary of lens results}
The total lens mass is determined to be $0.97 \pm 0.12 \, \mathrm{M}_{\odot}$. For the best-model mass ratio, this constrains the mass of the binary components as $M_{\mathrm{L},1} = 0.46 \pm 0.06  \, \mathrm{M}_{\odot}$ and $M_{\mathrm{L},2} = 0.52 \pm 0.06 \, \mathrm{M}_{\odot}$. The lens distance is determined to be $536 \pm 45 \, \mathrm{pc}$. This constrains the angular Einstein radius as $\theta_\mathrm{E} = 3.50 \pm 0.28 \, \mathrm{mas} $ and the physical Einstein radius is $R_\mathrm{E} = 1.87 \pm 0.13 \, \mathrm{AU} $. Given these results and assuming that the lensing stars are on the main sequence, we can conclude that the binary lens is consistent with late-K to early-M type stars.
For the lens distance determined, the recovered extinction is $A_{\mathrm{g},\mathrm{L}} = 0.26 \pm 0.02$ and $A_{\mathrm{i},\mathrm{L}} = 0.13 \pm 0.01$, such that the extinction-corrected magnitude of the combined binary lens components is $m_{\mathrm{g}_0,\mathrm{L}} = 17.52^{+0.43}_{-0.31}$ and $m_{i_0,\mathrm{L}}= 16.50^{+0.40}_{-0.29}$.

\subsection{Lens and source kinematics} 
The precise source parameters in \emph{Gaia} DR3 combined with the lens parameters determined above enable a full description of the kinematics of the lens and the source as shown in \citet{Rota21blx}. The photometric modelling constrains $t_\mathrm{E}$ in the geocentric frame, such that the geocentric relative proper motion is 
\begin{equation}
    \mu_{\mathrm{rel, geo}}(\mathrm{N,E}) = (-7.8 \pm 0.6, 16.6 \pm 1.3) \, \mathrm{mas\,yr^{-1}} \, .
\end{equation}
The heliocentric relative proper motion is given as
\begin{equation}
    \boldsymbol{\mu}_{\mathrm{rel, hel}} = \frac{\theta_\mathrm{E}}{t_\mathrm{E}}\hat{\boldsymbol{\pi}}_\mathrm{E} + \frac{\pi_{\mathrm{rel}}}{\mathrm{AU}} \boldsymbol{v}_{\oplus, \perp} \, \,,
\end{equation}
where $\boldsymbol{v}_{\oplus, \perp}$ is the Earth's projected velocity parallel to the lens plane, at the time $t_0$. It is determined from \emph{Astropy} ephemeris \citep{astropy} as $\boldsymbol{v}_{\oplus, \perp} (\mathrm{N,E})(t_0) =  (5.46, -28.17) \, \mathrm{km\,s^{-1}}$, such that the Earth's line of sight velocity to the microlensing event is $5.9 \, \mathrm{km\,s^{-1}}$.

The relative heliocentric proper motion is calculated to be 
\begin{equation}
    \boldsymbol{\mu}_{\mathrm{rel, hel}}(\mathrm{N,E}) = (-6.0 \pm 0.5, -7.4 \pm 1.0) \, \mathrm{mas\,yr^{-1}} \, .
\end{equation}
For the source proper motion (Table~\ref{tab:Gaia5param}), confirmed by TESS observations of $\mu_{\mathrm{RA}} = -2.331 \pm 0.027 \, \mathrm{mas\,yr^{-1}}$ and $\mu_{\mathrm{Dec}} = -2.762 \pm 0.038 \, \mathrm{mas\,yr^{-1}}$, the lens proper motion is determined to be 
\begin{equation}
    \boldsymbol{\mu}_{\mathrm{L, hel}}(\mathrm{N,E}) = (-3.3 \pm 0.5, 5.2 \pm 1.0) \, \mathrm{mas\,yr^{-1}} \, .
\end{equation}
The heliocentric lens proper motion is related to the heliocentric lens velocity via 
\begin{equation}
    v_{\mathrm{L, hel}} \, [\mathrm{km\,s^{-1}}] = 4.74 \times D_\mathrm{L} \, \mu_{\mathrm{L,hel}} \, [\mathrm{mas\,yr^{-1}}] \, ,
\end{equation}
where the 4.74 is a conversion factor from $\mathrm{mas}/\mathrm{yr}$ to $\mathrm{km}/\mathrm{s}$. It is calculated to be 
\begin{equation}
    \boldsymbol{v}_{\mathrm{L, hel}} (\mathrm{N,E}) = (-8.4 \pm 1.2, 13.2 \pm 2.9) \, \mathrm{km\,s^{-1}}\,\, ,
\end{equation}
which in Galactic coordinates equates to a lens velocity of 
\begin{equation}
    \boldsymbol{v}_{\mathrm{L}}(\mathrm{l,b}) = (12.9 \pm 2.2, 8.9 \pm 3.0) \, \mathrm{km\,s^{-1}} \,\, .
\end{equation}

\subsection{Keplerian orbital motion}
The Keplerian orbital-motion parameters of the best-fit model (Table~\ref{table:bestModel}) are tightly constrained, as evidenced by the MCMC chains in Fig.~\ref{fig:mcmcConv_2}. We therefore investigate the physical parameters of the binary lens system as described by \citet{Skowron2011} and \citet{BozzaVBMicrolensing}. Following their convention, the orbital parameters $(\gamma_1, \gamma_2, \gamma_3, r_s, a_s)$ are used to derive all elements of the bound elliptic orbital system and are listed in Table~\ref{tab:orbitalParams}. The separation of the binary lens objects at periastron, $r_{t_\mathrm{{peri}}} = a (1-e) \simeq 0.5 \, \mathrm{AU} $, and the kinematic vs potential check, $\beta_\perp = {E_{\mathrm{kin}, \perp}} \, / \, {|E_{\mathrm{pot}}|} \simeq 0.1 \leq 1.0 $, imply a bound and physical orbit.

\begin{table}[ht]
\caption{Lens orbital motion parameters and errors.}
\centering
\label{tab:orbitalParams}
\renewcommand{\arraystretch}{1.3}
\begin{tabular}{lr}
\hline
\textbf{Orbital Parameters} &  \\
\hline
    Reference time $t_{\mathrm{ref}}$ & $2459406.3$ \\
    Orbital period $P \,[\mathrm{yr}]$  & $0.67^{+0.04}_{-0.04}$ \\
    Semi-major axis $a \, [\mathrm{AU}]$ & $0.763^{+0.016}_{-0.015}$ \\
    Eccentricity $e$  & $0.30^{+0.03}_{-0.03}$ \\
    Inclination $i \, [\mathrm{deg}]$ & $116.1^{+0.5}_{-0.5}$ \\
    Arg. of periastron $\omega \, [\mathrm{deg}]$ & $12.9^{+0.8}_{-0.8}$ \\
    Long. asc. node $\Omega \, [\mathrm{deg}]$ & $169.3^{+0.5}_{-0.5}$ \\
    Periastron epoch $t_{\mathrm{peri}} \, [\mathrm{JD}]$ & $2459523.6^{+7.0}_{-8.0}$ \\
    RV semi-amplitude $K_1 \, [\mathrm{km\,s^{-1}}]$ & $16.9^{+0.9}_{-0.9}$ \\
\hline
\end{tabular}
\end{table}

\subsection{Follow-up observations}
As the lens of the microlensing event is bright and the relative proper motion of the source and lens has been determined, we are able to determine when the source and lens will have an angular separation large enough to conduct follow-up observations of the binary lens. Radial-velocity follow-up has been used before to test microlensing models for binary lenses with orbital motion. The microlensing event OGLE-2009-BLG-020 \citep{Skowron2011} estimated the full orbital parameters of a binary lens, which was followed up and confirmed by \citet{Yee2016}. The microlensing event OGLE-2011-BLG-0417 \citep{Shin_2012} on the other hand was challenged by RV observations \citep{Boisse_2015} and high-resolution imaging \citep{Santerne_2016}, which resulted in a refined model published in \citet{Bachelet2018OGLE0417}.

Due to the large difference in brightness of the source and the lens in the Gaia20fnr microlensing event, a seeing-limited high-resolution observation would require a separation of $\theta \gtrsim 1^{\prime\prime}$ to measure the radial-velocity signal of the binary system. A diffraction-limited or AO-assisted observation would require $\theta \gtrsim 0.1^{\prime\prime}$. With the estimated relative proper motion values of the source and lens, these observations would be feasible $\sim10.5$ and $\sim105$ years after $t_0$ for the diffraction-limited and seeing-limited observations, respectively. Although radial-velocity and high-resolution imaging observations will also require several decades before the source and lens are cleanly distinguishable, they could ultimately provide an independent confirmation of the system geometry and mass ratio of the lens.

\subsection{\emph{Gaia} astrometry}
\begin{figure}
    \centering
    \includegraphics[width=\columnwidth]{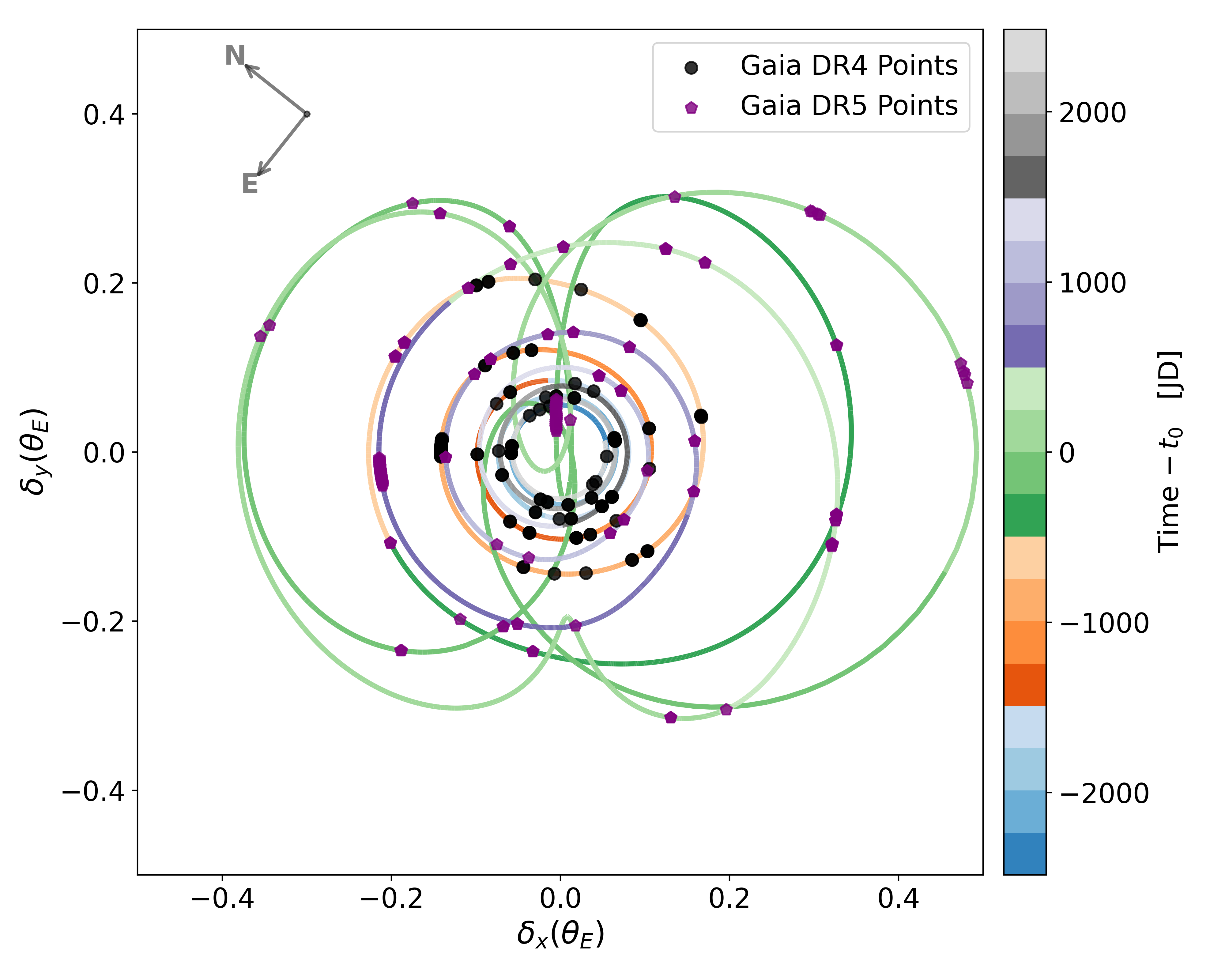}
    \caption{Predicted centroid motion of the source of the Gaia20fnr microlensing event. The points marked as \emph{Gaia} data were taken at timestamps determined with the \emph{Gaia} Observation Forecast Tool (GOST)}
    \label{fig:astrometricSignal}
\end{figure}

\emph{Gaia} DR4 will publish astrometric time series for all sources, which, for microlensing events brighter than about $G \lesssim 16 \, \mathrm{mag}$ will be of sufficient accuracy to detect the astrometric microlensing signal \citep{Rybicki2018}. The strength of using \emph{Gaia} astrometry to test microlensing models has been demonstrated for the bright \emph{Gaia} DR3 event GaiaDR3-ULENS-001, for which the anomalous DR3 astrometric solution can be explained by the astrometric microlensing signal implied by the photometric model \citep{JablonskaUlens001, WyrzykowskiDR3Microlensing}. For the source of the microlensing event analysed in this study, which has an even brighter baseline magnitude of $m_\mathrm{G} = 13.11 \, \mathrm{mag}$, the astrometric data will therefore allow a direct measurement of $\theta_\mathrm{E}$. Although the full astrometric signal that overlaps with the entire photometric event will not be published in \emph{Gaia} DR4, the astrometric microlensing signal is observable well before and well after the photometric microlensing signal and could therefore be detectable with earlier data \citep{Proft_2011, Belokurov_2002}. The astrometric time-series data released in DR4 will be from observations during the first 66 months of the mission, and could therefore already contain a detectable astrometric microlensing signal. We simulated the predicted astrometric signal based on the microlensing parameters from the photometric model. Together with the timestamps of predicted \emph{Gaia} observations, it is shown in Fig.~\ref{fig:astrometricSignal}\footnote{GOST: \url{https://Gaia.esac.esa.int/gost/}}.

% In Conclusion include discussion of light from Swift:
%  - Normal optical obseravtions (\emph{Gaia} and LCO) have flux ratio of ~0.98
%  - SWIFT UV(UVW1 and UVW2) observations fit the blend flux consistent with zero, indicating that no light is coming from the blend/lens.
% - NEOWISE(W1) observations fit 0.985 flux ratio
% - NEOWISE(W2) observations fit 0.970 flux ratio
% - Optical Swift is also around 0.93 to 0.97 and in agreement with the bands.
% Ultraviolet (UV) light is dominated by hotter stars. 
% Our source star is an early K-type star.
% Our lens is a late K-type star and an early M-type star.
% We would expect our source to emit more UV light than the blend (ie higher SWIFT flux ratio).
% We would expect strong IR signals from both so a signal around that of optical makes sense.
% Discuss SED.

\section{Conclusions}
We analysed the triple-peaked, non-caustic-crossing binary microlensing event Gaia20fnr, whose complex morphology required a full Keplerian orbital-motion model to reproduce the photometric data. The light curve contains data from four space telescopes (\emph{Gaia}, NEOWISE, Swift, and TESS) as well as ground-based telescopes around the world coordinated through the BHTOM network. The photometric data was complemented by spectroscopic observations and by parameters from \emph{Gaia} DR3 to set constraints on the source.

By incorporating information about the lens from the blend-flux in the best-fit microlensing model, we found that Gaia20fnr is best explained by a K2-type giant source at a distance of 3.1 kpc lensed by a binary system at 0.5 kpc. The lens consists of an early-M star ($M_{\mathrm{L,1}} = 0.46 \pm 0.06 \, \mathrm{M_\odot}$) and a late-K star ($M_{\mathrm{L,2}} = 0.52 \pm 0.06 \, \mathrm{M_\odot}$), orbiting each other with a period of $0.67 \pm 0.04$ years and a semi-major axis of $0.76 \pm 0.02\,\mathrm{AU}$. The Swift UV observations show a larger ratio of the measured flux coming from the source than for optical and IR observations. This is expected for the derived stellar parameters of the K2-type giant source star and the lens, as UV emission is dominated by hotter stars. The kinematics of the microlensing event predicts significant relative proper motion of the source and the lens and a measurable radial-velocity semi-amplitude. The main results are summarised in Table~\ref{tab:summarizedResults}.

Gaia20fnr is one of only a handful of microlensing events for which a full Keplerian orbital solution has been derived and for which the results will be verifiable by follow-up observations. Upcoming \emph{Gaia} DR4 and DR5 astrometric time-series data, together with future radial-velocity monitoring and high-resolution imaging, will uniquely verify both the source classification and the lens-system architecture.

With the onset of deeper, high-cadence, and wide-area microlensing surveys such as the Vera C. Rubin Observatory’s Legacy Survey of Space and Time (LSST) \citep{LSSTSajadian} and the Nancy Grace Roman Space Telescope \citep{RomanPenny}, the discovery rate of long, complex, and parallax- or orbital-motion-dominated microlensing events will increase dramatically. Events like Gaia20fnr demonstrate that densely sampled, multi-mission photometry combined with source constraints from spectroscopy and the \emph{Gaia} DR3 catalogue enable precise characterisation of nearby binary lenses, including complete orbital solutions. Additionally, future space-based missions like the proposed \emph{GaiaNIR} mission \citep{GaiaNIR}, which will be the next generation astrometric survey providing unique and ultra-precise astrometric time-series, could provide the precision necessary to break common microlensing degeneracies. This will thereby enable secure physical characterisations of complex microlensing events such as the one presented in this study.

\begin{table}[ht]
\caption{Summary of the key parameters describing the Gaia20fnr microlensing event}
\centering
\label{tab:summarizedResults}
\renewcommand{\arraystretch}{1.3}
\begin{tabular}{lr}
\hline
\textbf{Event Parameters} &  \\
\hline
    $D_\mathrm{S} \, [\mathrm{pc}]$ & $ 3109 \pm 105$ \\
    $D_\mathrm{L} \, [\mathrm{pc}]$ & $ 536 \pm 45$ \\
    $\theta_\mathrm{E} \, [\mathrm{mas}]$ & $ 3.50 \pm 0.28$ \\
    % $\mu_{rel, geo} \, [\mathrm{mas\,yr^{-1}}]$ & $ 18.4 \pm 1.5$ \\
    $\mu_{\mathrm{rel, geo}}(\mathrm{N,E}) \, [\mathrm{mas\,yr^{-1}}]$ & $(-7.8 \pm 0.6, 16.6 \pm 1.3)$ \\
    % $\mu_{S, hel} \, [\mathrm{mas\,yr^{-1}}]$ & $ 3.53 \pm 0.01$ \\
    $\mu_{\mathrm{S, hel}}(\mathrm{N,E}) \, [\mathrm{mas\,yr^{-1}}]$ & $(2.72 \pm 0.01, -2.25 \pm 0.01)$ \\
    % $\mu_{L, hel} \, [\mathrm{mas\,yr^{-1}}]$ & $ 6.2 \pm 1.1$ \\
    $\mu_{\mathrm{L, hel}}(\mathrm{N,E}) \, [\mathrm{mas\,yr^{-1}}]$ & $(-3.3 \pm 0.5, 5.2 \pm 1.0)$ \\
    $M_{\mathrm{L},1} \, [\mathrm{M_\odot}]$ & $ 0.46 \pm 0.06$ \\
    $M_{\mathrm{L},2} \, [\mathrm{M_\odot}]$ & $ 0.52 \pm 0.06$ \\
    $P \, [\mathrm{yr}]$ & $0.67 \pm 0.04$ \\
    $K_1 \, [\mathrm{km\,s^{-1}}]$ & $16.9 \pm 0.9$ \\
\hline
\end{tabular}
\end{table}

\begin{acknowledgements}
BHTOM.space is based on the open-source TOM Toolkit by LCO and has been developed with funding from the OPTICON-RadioNet Pilot (ORP) of the European Union's Horizon 2020 research and innovation programme under grant agreement No 101004719 (2021-2025). This project has received funding from the European Union's Horizon Europe Research and Innovation programme ACME under grant agreement No 101131928 (2024-2028). {\L}.W. acknowledges support from the Polish National Science Centre DAINA grant No 2024/52/L/ST9/00210. E.S., M.M., J.Z., E.P., V.C., and U.J. acknowledge funding from the Research Council of Lithuania (LMTLT, grant No. S-LL-24-1). 

Support for this project is provided by ANID's Millennium Science Initiative through grant ICN12\_009, awarded to the Millennium Institute of Astrophysics (MAS), and by ANID's Basal project FB210003.
YT acknowledges the support of DFG priority program SPP 1992 “Exploring the Diversity of Extrasolar Planets” (TS 356/3-1).
RAS and EB gratefully acknowledge support from the NASA XRP Program, through grant number 80NSSC19K0291.  This work was authored by employees of Caltech/IPAC under Contract No. 80GSFC21R0032 with the National Aeronautics and Space Administration.
This project has received funding from the European Union's Horizon 2020 research and innovation program under grant agreement No. 101004719 (OPTICON - RadioNet Pilot). 
This work is supported by the Polish MNiSW grant DIR/WK/2018/12.
Partial support for MR is provided by the Direcci{\'o}n de Investigaci{\'o}n of the Universidad Cat{\'o}lica de la Sant{\'i}sima Concepci{\'o}n with the project DIREG 10/2023 and by ANID with the project QUIMAL230011.

KK greatetely acknowledges the support from the NSF AAG program, through grant number 2206828.

The research of JMe was supported by the Czech Science Foundation (GACR) project no. 24-10608O.

This work is partially supported by the Fundamental Fund of Thailand Science Research and Innovation (TSRI) through the National Astronomical Research Institute of Thailand (Public Organization) (FFB690078/0269).

We thank the Neil Gehrels \textit{Swift} Observatory team for promptly approving and executing our Target of Opportunity observations. 
M.W. thanks Mateusz Jan Mr\'{o}z and the Microlensing Journal Club at OAUW for helpful discussions regarding the modelling of the light curve.

\end{acknowledgements}
   
% \end{thebibliography}
\bibliographystyle{aa}
\bibliography{bibs}

\begin{appendix} % appendix
\section{Photometric data and follow-up}
Table~\ref{tab:telescopeInfo} contains all the information about the telescopes and instruments involved in the ground-based follow-up observations of the Gaia20fnr microlensing event. The data for the follow-up observations were processed by the BHTOM tool \citep{BHTOM1}. Table~\ref{tab:allData} contains information on the observations made with all telescopes, including the ground-based follow-up observations, surveys, and space telescopes, in connection with the Gaia20fnr event. 

% #######################################
% Table 1: Observatories/Telescopes
% #######################################

\begin{table*}[htbp]
\caption{Instruments involved with ground-based follow-up observations of the Gaia20fnr microlensing event.}
\label{tab:telescopeInfo}
\centering
\begin{tabularx}{\textwidth}{
    l
    >{\hsize=9\hsize\arraybackslash}X
    >{\hsize=6\hsize\arraybackslash}X
    l l l l
}
\hline\hline
\textbf{Telescope Code} & \textbf{Telescope/Observatory Name} & \textbf{Location} & \textbf{Long. [deg]} & \textbf{Lat. [deg]} & \textbf{Alt. [m]} & \textbf{Reference} \\
\hline
Abastumani-150 & Georgian National Astrophysical Observatory & Abastumani, Georgia & 42.8 & 41.8 & 1650 & - \\
Adiyaman-60 & Adiyaman Observatory & Adiyaman, Turkey & 38.2 & 37.8 & 600 & - \\
Danish-154 & La Silla Observatory & La Silla, Chile & -70.7 & -29.3 & 2375 & \citet{Andersen1995} \\
Flarestar-25 & Flarestar Observatory & San Gwann, Malta & 14.5 & 35.9 & 126 & - \\
GoChile-40 & El Sauce Observatory & Río Hurtado Valley, Chile & -70.8 & -30.5 & 1560 & \citet{ElSauceObservatory} \\
LCO-cpt-100 & LCO - South African Astronomical Observatory & Sutherland, South Africa & 20.8 & -32.4 & 1760 & \citet{LCOBrown} \\
LCO-lsc-100 & LCO - Cerro Tololo Inter-American Observatory & Cerro Tololo, Chile & -70.8 & -30.2 & 2201 & \citet{LCOBrown} \\
LCO-coj-100 & LCO - Siding Spring Observatory & Siding Spring, Australia & 149.1 & -31.3 & 1165 & \citet{LCOBrown} \\
LCO-elp-100 & LCO - McDonald Observatory & Jeff Davis County, Texas & -104.0 & 30.7 & 2077 & \citet{LCOBrown} \\
LCO-tfn-100 & LCO - Teide Observatory & Tenerife & -16.5 & 28.3 & 2390 & \citet{LCOBrown} \\
PROMPT-MO-1-40 & Meckering Observatory & Meckering, Australia & 117.0 & -31.6 & 650 & \citet{PromptObservatory} \\
PROMPT5-41 & Cerro Tololo Inter-American Observatory & Cerro Tololo, Chile & -70.8 & -30.2 & 2286 & \citet{PromptObservatory} \\
PROMPT6-40 & Cerro Tololo Inter-American Observatory & Cerro Tololo, Chile & -70.8 & -30.2 & 2286 & \citet{PromptObservatory} \\
REM-60 & La Silla Observatory & La Silla, Chile & -70.7 & -29.3 & 2400 & \citet{REMobservatory} \\
ROAD-40 & Remote Observatory Atacama Desert & San Pedro de Atacama, Chile & -68.2 & -23.0 & 2500 & \citet{RoadObservatory} \\
RRRT-60 & Fan Mountains Observatory & Virginia, USA & -78.7 & 37.9 & 1683 & \citet{RRRTTelescope} \\
SUTO-Otivar-30 & Silesian University of Technology Observatories & Granada, Spain & -3.7 & 36.8 & 200 & - \\
TRT-SBO-70 & Thai Robotic Telescopes, Spring Brook Observatory & Springbrook, Queensland, Australia & 149.1 & -31.3 & 1020 & - \\
Terskol-60 & Terskol Observatory & Terskol, North Caucasus & 42.5 & 43.3 & 3143 & \citet{TerskolObservatory} \\
Terskol-200 & Terskol Observatory & Terskol, North Caucasus & 42.5 & 43.3 & 3143 & \citet{TerskolObservatory} \\
UZPW-50 & University of Zielona Góra PlaneWave L500 & Badajoz, Spain & -6.6 & 38.2 & 560 & \citet{ROTUZ} \\
Warrumbungle-51 & Warrumbungle Observatory & Siding Spring, Australia & 149.2 & -31.3 & 552 & - \\
ZAO-20 & Znith Astronomy Observatory & Malta & 14.4 & 35.9 & 113 & - \\
\hline
\end{tabularx}
\end{table*}

\newpage

\begin{table*}[htbp]
\caption{Telescope-specific information showing the time range and number of data points in each filter for observations of the Gaia20fnr microlensing event.}
\label{tab:allData}
\centering
\begin{tabularx}{\textwidth}{l l l l}
% \hline\hline
\hline
\textbf{Telescope code} & \textbf{First epoch [MJD]} & \textbf{Last epoch [MJD]} & \textbf{Number of Points (Filter)} \\
\hline
Abastumani-150 & 59510.54 & 59527.48 & 13(V),12(r),6(I),2(i) \\
Adiyaman-60 & 59256.32 & 59319.20 & 4(g),2(I),2(R),2(r),1(z) \\
Danish-154 & 59533.65 & 59867.73 & 8(R),8(V),7(I),6(B) \\
Flarestar-25 & 59542.45 & 59636.27 & 11(V),8(I) \\
GoChile-40 & 60025.56 & 60029.55 & 9(g),6(V),6(r) \\
LCO-cpt-100 & 59206.83 & 60029.73 & 415(i),306(g),3(B),1(I),1(R) \\
LCO-lsc-100 & 59209.21 & 60034.98 & 442(i),324(g),6(B),1(I),1(U),1(u) \\
LCO-coj-100 & 59205.56 & 60027.40 & 252(i),167(g) \\
LCO-elp-100 & 59215.24 & 59849.49 & 19(i),13(g) \\
LCO-tfn-100 & 59472.24 & 59576.00 & 27(i),16(g) \\
PROMPT-MO-1-40 & 59288.97 & 59770.41 & 3(V),2(I),2(r),1(i) \\
PROMPT5-41 & 59680.49 & 59905.85 & 16(g),15(i),15(r),4(V),2(I),1(R) \\
PROMPT6-40 & 59277.62 & 59296.61 & 33(V),29(I),21(R),12(r),3(i) \\
REM-60 & 59243.57 & 60035.57 & 1014(i),932(r),762(z),616(g),536(V),398(R),245(I),4(B),1(u) \\
ROAD-40 & 59233.55 & 59646.53 & 2346(V),2188(B),2121(I),81(R),3(U) \\
RRRT-60 & 59843.92 & 59861.92 & 2(g),2(i),2(r) \\
SUTO-Otivar-30 & 59828.68 & 60037.34 & 66(B),57(V),57(r),45(I),14(R),10(i),2(g) \\
TRT-SBO-70 & 59221.95 & 59471.18 & 64(V),41(I),15(R),3(r),1(i) \\
Terskol-60 & 59621.33 & 59635.29 & 8(g),7(R),7(V),6(I),1(z) \\
Terskol-200 & 59465.55 & 59629.31 & 24(V),18(R),17(g),15(I),13(r),8(i),7(B),7(z),1(u) \\
UZPW-50 & 59908.71 & 60043.48 & 147(g),145(r),133(i),14(V),11(B),9(R) \\
Warrumbungle-51 & 59249.03 & 59260.06 & 1(B),1(R),1(V),1(r) \\
ZAO-20 & 59579.48 & 59649.37 & 58(I),44(V) \\
\hline
ASAS-SN & 56593.58 & 60713.11 &  3022(g),852(V) \\
ATLAS & 57297.56 & 60347.92 & 2563(o),670(c),4(g),3(r),1(H) \\
ZTF & 58439.27 & 60286.35 & 90(zr),80(zg) \\
\hline
\emph{Gaia} & 57069.56 & 60325.48 & 255(G) \\
NEOWISE & 56734.61 & 60369.00 & 354(W2),353(W1) \\
Swift & 59327.64 & 59657.45 & 17(B),17(U),17(UVW1),16(UVW2),16(V) \\
TESS & 58467.79 & 60689.44 & 15633(TESS) \\
\end{tabularx}
\end{table*}

\clearpage

\section{Fit Convergence}
\begin{figure*}
\centering
\includegraphics[width=0.6\textwidth]{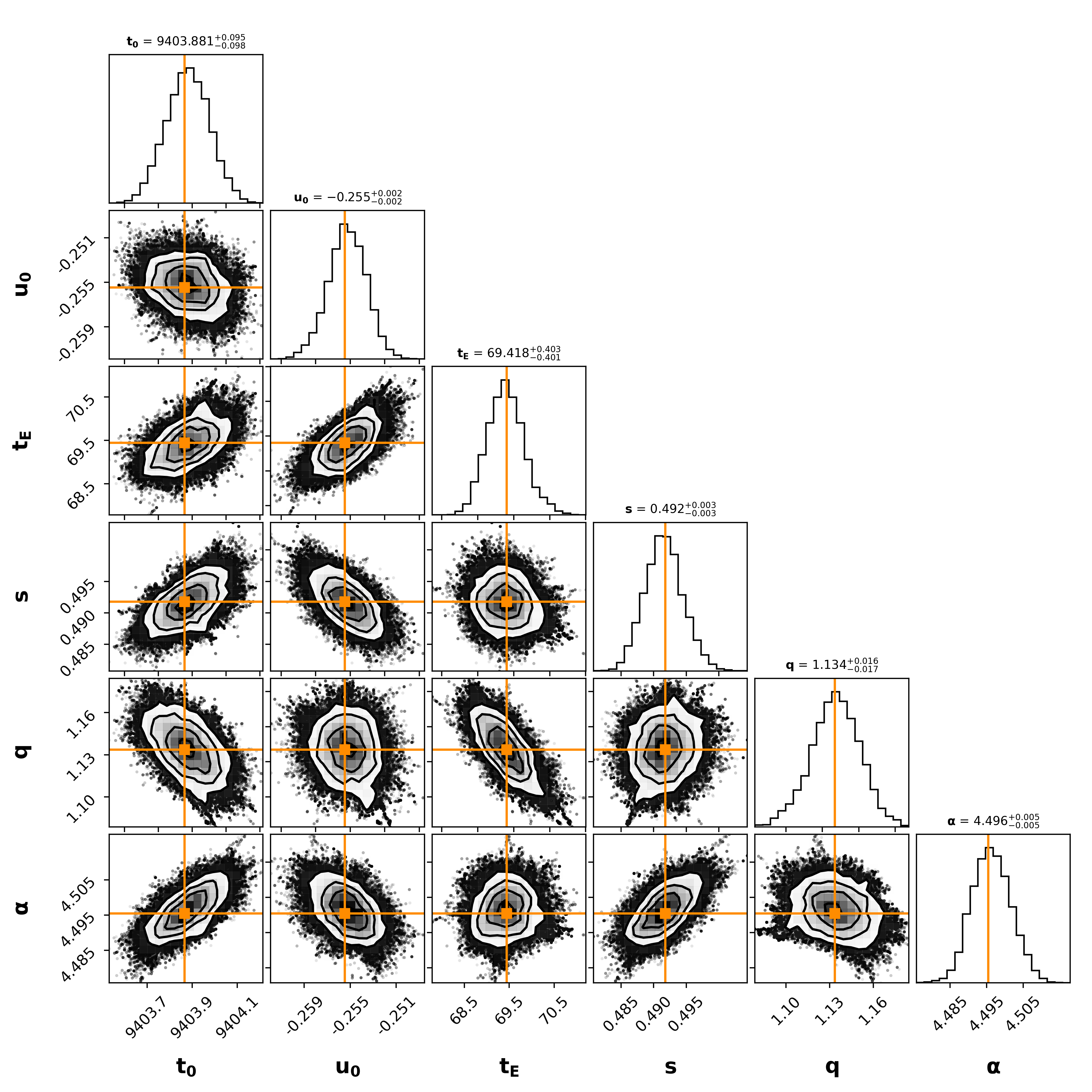}
\includegraphics[width=0.6\textwidth]{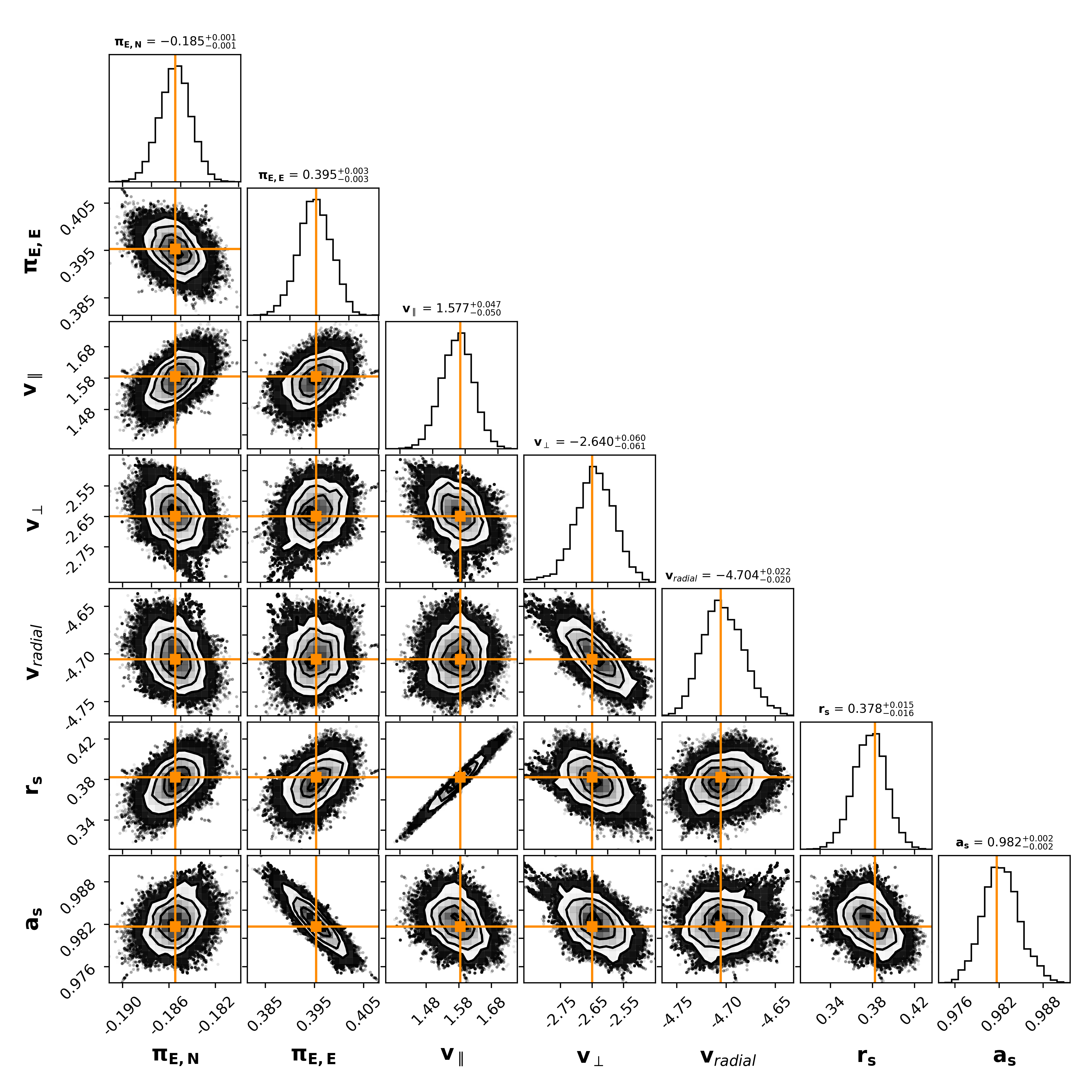}
\caption{Confidence contours in $\chi^{2}$ space for the MCMC-derived binary-lens microlensing parameters (top)  and microlens parallax and orbital motion parameters (bottom) of the best model including orbital motion. The $1\sigma$, $2\sigma$, and $3\sigma$ regions are shown in black, dark grey, and light grey shading. Samples lying beyond the $3\sigma$ boundary are plotted as black points. The best-model parameters (see Table~\ref{table:bestModel}) are indicated by the orange lines and squares. The visualization was produced with the \texttt{corner} package \citep{DFM_Corner}.}
\label{fig:mcmcConv_2}
\end{figure*}

Fig.~\ref{fig:mcmcConv_2} contains the confidence contours of MCMC-derived binary-lens microlensing parameters and the confidence contours of MCMC-derived parallax and orbital motion parameters for the same model and MCMC analysis.
\clearpage

\end{appendix}

\end{document}